\documentclass[%
 reprint,
superscriptaddress,
 amsmath,amssymb,
 aps,
]{revtex4-1}

\usepackage{graphicx}
\usepackage{dcolumn}
\usepackage{bm}
\usepackage{physics}
\usepackage{multirow}
\usepackage{amsthm}
\usepackage{apptools}
\usepackage{chngcntr}
\usepackage{mathtools}
\usepackage{subcaption}
\usepackage{color}
\AtAppendix{\counterwithin{theorem}{section}}
\makeatletter
\def\blfootnote{\xdef\@thefnmark{}\@footnotetext}
\makeatother

\begin{document}

\preprint{APS/123-QED}

\title{2-D Compass Codes}

\author{Muyuan Li}
\affiliation{School of Computational Science and Engineering, Georgia Institute of Technology, Atlanta, GA 30332, USA}
\author{Daniel Miller} 
\affiliation{Institut f\"ur Theoretische Physik III\hspace{.5mm}, Heinrich-Heine-Universit\"at D\"usseldorf, D-40225 D\"usseldorf, Germany}
\author{Michael Newman}
\email{michael.newman@duke.edu}
\affiliation{Departments of Electrical and Computer Engineering, Chemistry, and Physics, Duke University, Durham, NC, 27708, USA}
\author{Yukai Wu}
\affiliation{ Department of Physics,
  University of Michigan, Ann Arbor, MI 48109, USA}
\author{Kenneth R. Brown}
\email{kenneth.r.brown@duke.edu}
\affiliation{Departments of Electrical and Computer Engineering, Chemistry, and Physics, Duke University, Durham, NC, 27708, USA}

\date{\today}
\begin{abstract}
The compass model on a square lattice provides a natural template for building subsystem stabilizer codes. The surface code and the Bacon-Shor code represent two extremes of possible codes depending on how many gauge qubits are fixed.  We explore threshold behavior in this broad class of local codes by trading locality for asymmetry and gauge degrees of freedom for stabilizer syndrome information.  We analyze these codes with asymmetric and spatially inhomogeneous Pauli noise in the code capacity and phenomenological models.  In these idealized settings, we observe considerably higher thresholds against asymmetric noise.  At the circuit level, these codes inherit the bare-ancilla fault-tolerance of the Bacon-Shor code.
\end{abstract}

\maketitle


\section{Introduction}

At the heart of scalable quantum computing is fault-tolerance.  The celebrated quantum threshold theorem \cite{Aliferis:2006, Knill:1996b, Aharonov:1997} ensures that with sufficiently accurate components, we can perform arbitrarily long quantum computations with polylogarithmic overhead.  For physical systems that prefer local interactions, topological codes have emerged as leading candidates for fault-tolerant quantum computation \cite{Dennis:2002, Bombin:2013, Tomita:2014, Yoder:2017b, Fowler:2012, terhal2015quantum, o2017density,trout2018simulating}.  Among these, the surface code is a particularly enticing choice, offering depolarization accuracy thresholds in excess of $15\%$ assuming noiseless error-correction with a planar architecture \cite{Wang:2009}.

Another code family which has generated significant interest are the subsystem Bacon-Shor codes \cite{Bacon:2006}.  These codes have many desirable properties: their gauge group is $2$-local, measurements can be performed with bare ancilla with virtually no loss in performance \cite{Aliferis:2007, Li:2018}, and they support fault-tolerance schemes that avoid costly magic-state distillation \cite{Yoder:2017}.  Unfortunately, while Bacon-Shor codes offer some of the highest concatenated thresholds \cite{Aliferis:2007}, they fail to have any threshold when grown as a local family on a lattice without concatenation \cite{Pastawski:2009}.

In the present article, we investigate codes derived from the quantum compass model on a square lattice \cite{nussinov2015compass}.  This model provides a natural framework for constructing subsystem and subspace stabilizer codes.  These codes can be viewed as different gauge-fixes of the Bacon-Shor code, and so include the (rotated) surface codes, as well as codes with certain topological defects, as members \cite{Tomita:2014,Yoder:2017b}.  While we focus on a subfamily of (generalized) surface codes \cite{delfosse2016generalized} with desirable fault-tolerance properties, the design space for these codes is much larger.  Two advantages of this family are its malleability, making it suitable for correcting asymmetric noise, and fault-tolerant bare-ancilla syndrome extraction inherited from measuring along the gauges of the template Bacon-Shor code.

Tailoring codes and decoders to specific noise models can often yield fruitful improvements in threshold scaling.  For biased noise models, one can choose fault-tolerance schemes and gates that take advantage of asymmetric error rates \cite{PhysRevA.78.052331, PhysRevA.92.062309, stephens2013high}.  Indeed, simply choosing the right decoder can yield tremendous gains in the effective threshold \cite{Tuckett:2018,delfosse2014decoding,nickerson2017analysing,darmawan2018efficient}.  One can even customize codes directly to device level noise \cite{PhysRevApplied.8.064004} or biased error-rates \cite{Napp:2012, PhysRevA.87.032310}.  Such asymmetric noise models are motivated experimentally by the observation that dephasing noise dominates certain quantum computing architectures \cite{aliferis2009fault}.  By modifying the stabilizers and boundaries of a planar code directly, one can also obtain denser packings of logical qubits \cite{delfosse2016generalized} and optimized performance with respect to erasures \cite{delfosse2016linear}.

We similarly modify the geometry of planar codes using the convenient language of compass codes, adapting the density of the syndrome information to better correct biased and spatially dependent Pauli noise.  To quantify the value of this adapted syndrome data, we consider randomized \cite{Dennis:2002,Bombin:2010b}, minimum-weight perfect matching \cite{Dennis:2002,Wang:2009}, and union-find \cite{delfosse2017almost} decoders that treat $X$- and $Z$-type errors independently. We choose different decoders depending on the context, but generally observe similar performance across all three.  In particular, we expect that tuning correlated decoders to account for these different noise models will boost code performance even further \cite{Tuckett:2018}.

The idea is simple: one should tesselate a lattice according to the relative likelihood of errors in that part of the lattice. Although these codes remain local, there is a trade-off between the locality of their stabilizers and their robustness against asymmetric noise, similar to \cite{stephens2013high}.  We analyze these codes numerically in the code capacity and phenomenological noise models, and observe considerably higher thresholds against asymmetric noise in these idealized settings.  We leave a discussion of the challenges posed by circuit-level noise to the conclusion.

The paper is structured as follows.  In Section \ref{Background}, we introduce $2$-D compass codes and the noise models we consider.  In Section \ref{Threshold Scaling}, we determine the threshold behavior in two randomized families of codes interpolating between Bacon-Shor codes, surface codes, and Shor's code.  In Section \ref{Asymmetric Noise}, we quantify the threshold of $2$-D compass codes tailored for different asymmetric noise models.  In Section \ref{ft}, we demonstrate fault-tolerance for the compass code family using only bare-ancilla syndrome extraction.  We conclude with some discussion in Section \ref{Discussion}.

\section{Background}\label{Background}

\subsection{$2$-D Compass Codes}

The quantum compass model on a square lattice is defined generally by the Hamiltonian \cite{dorier2005quantum, kugel1973crystal}, $$H =\sum\limits_i\sum_{j \neq L-1} J_X X_{i,j} X_{i,j+1} + \sum_{i \neq L-1} \sum_{j} J_Z Z_{i,j}Z_{i+1,j}.$$ Here, $(i,j)$ indexes a qubit according to its displacement from the top-left corner of the lattice.  Closely connected with this model are Bacon-Shor codes, which are stabilizer subsystem codes with gauge operators realized by the two-body interaction terms of the compass model \cite{Bacon:2006}.  This family is a standard example of codes requiring only local measurements for error-correction, but which are \emph{not} topological, with stabilizers that extend the length of the lattice.

The gauge group of a Bacon-Shor code is generated by $\mathcal{G} = \langle X_{i,j} X_{i,j+1}, Z_{i,j}Z_{i+1,j}\rangle$, with stabilizer group $\mathcal{S} = \langle \prod_j X_{i,j}X_{i+1,j}, \prod_i Z_{i,j}Z_{i,j+1}\rangle$.  When defined on an $L \times L$ lattice, these $2L$-body stabilizer generators leave us with $(L-1)^2$ gauge degrees of freedom to format as we please.

Our tool for constructing compass codes will be the method of gauge-fixing, by which we can insert gauge transformations into the stabilizer group \cite{Bombin:2013, Paetznick:2013}.  Operationally, this corresponds to inserting a gauge operator $g$ into $\mathcal{S}$ and then removing the set of all gauge operators $h$ which anticommute with $g$ from $\mathcal{G}$.  Note that as Bacon-Shor codes are CSS codes \cite{calderbank1996good, steane1996error}, if we perform fixes of either $X$- or $Z$-type, we will preserve the CSS structure.

\begin{figure}[htb!]
\captionsetup[subfigure]{labelformat=empty}
        \begin{subfigure}[b]{0.24\textwidth}
                \centering
                \includegraphics[width=0.95\linewidth]{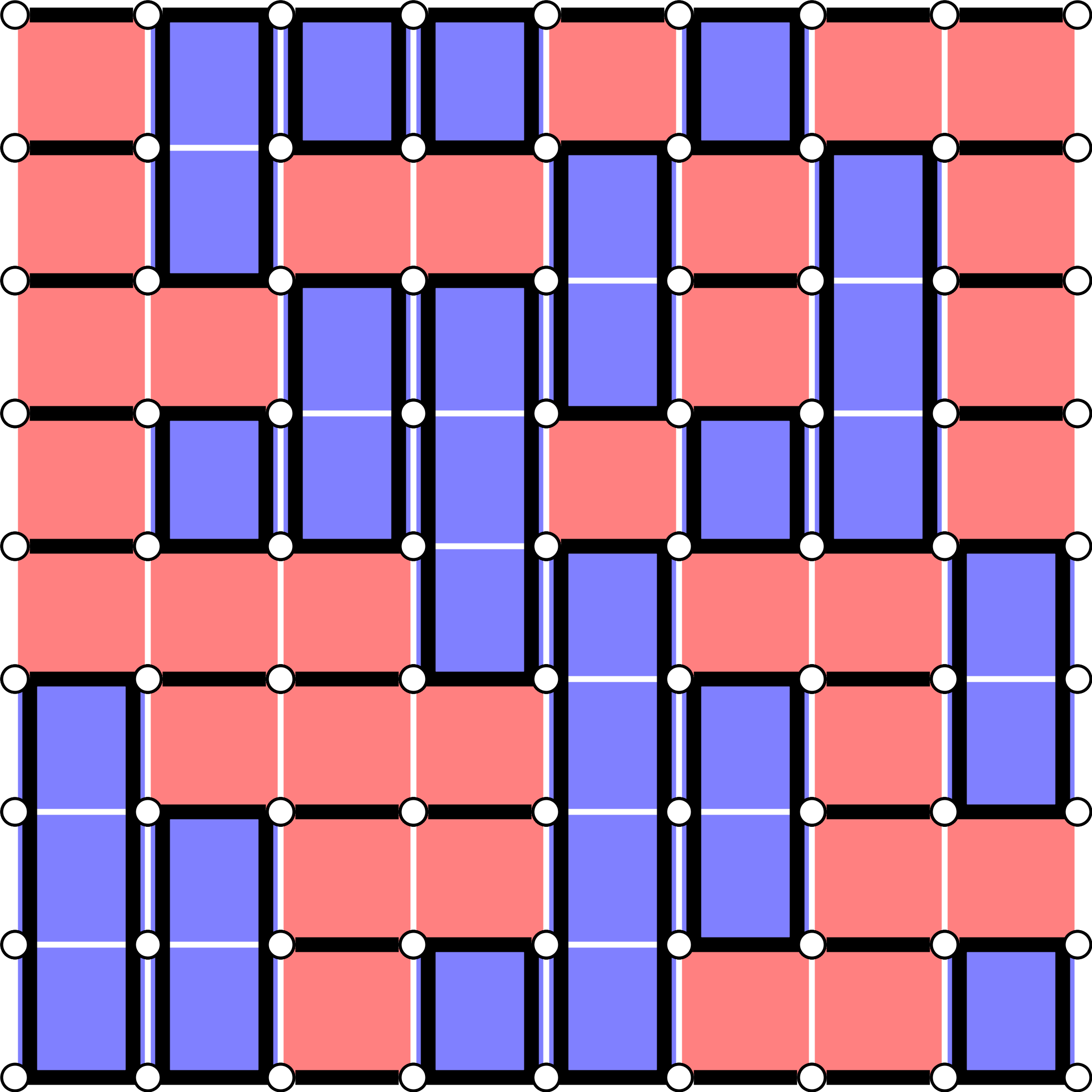}
        \end{subfigure}%
        \begin{subfigure}[b]{0.24\textwidth}
                \centering
                \includegraphics[width=0.95\linewidth]{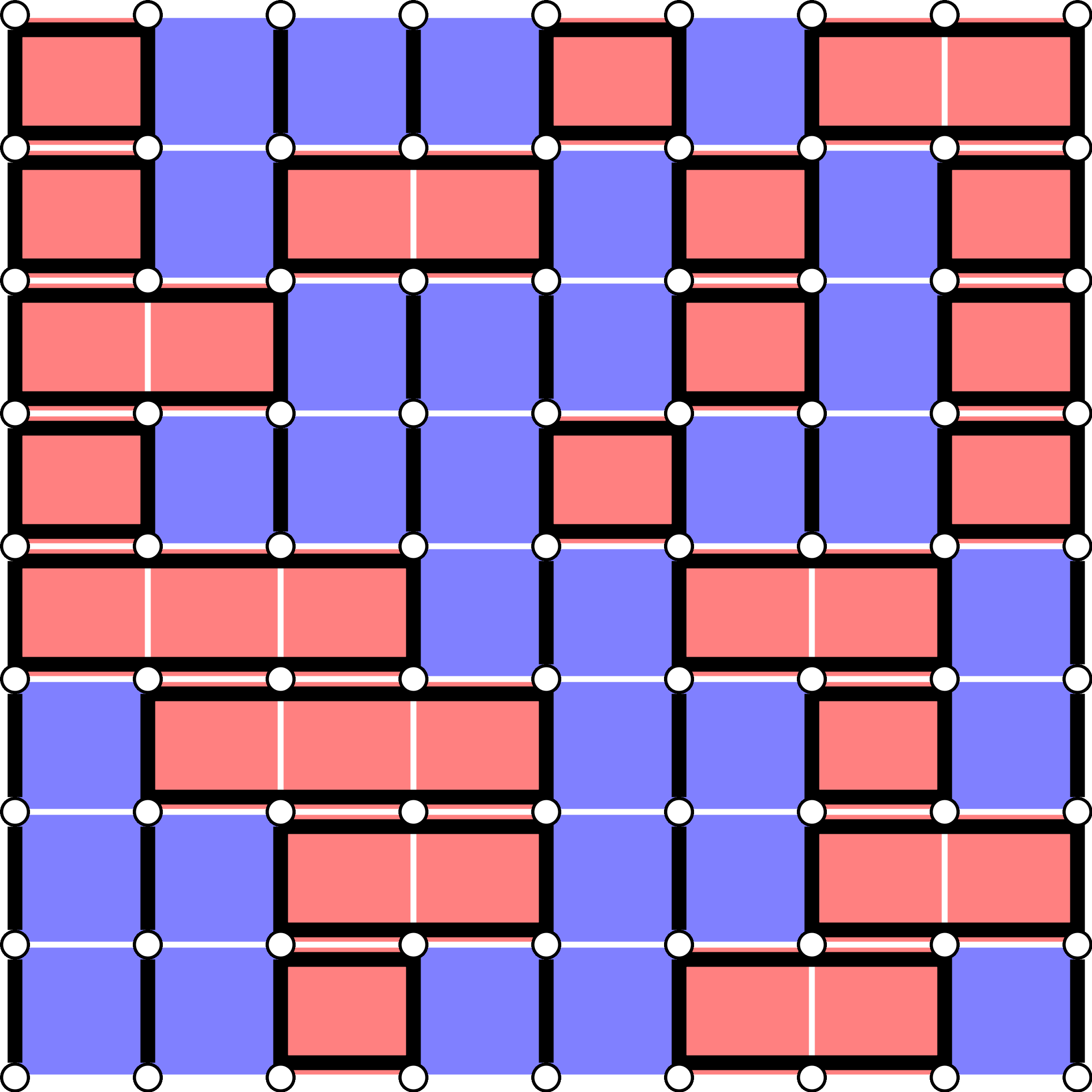}
        \end{subfigure}
        \caption{An example of a compass code on a $9\times9$ lattice.  Red and blue plaquettes represent cuts in the $Z$-type and $X$-type stabilizers, respectively. The bold lines outline the $Z$- and $X$-type stabilizers in the left- and right-side pictures, respectively.  As there are no blank plaquettes, all of the gauge degrees of freedom are fixed.}
        \label{cut_codes}
\end{figure}

We focus on a subclass of surface codes that are easy to specify via a coloring of the lattice, see Figure \ref{cut_codes}.  In that graphical language, red plaquettes correspond to ``cuts'' in the vertical $Z$-type stabilizers.  We index plaquettes according to the index of their top-left qubit; then, for a red plaquette in the $(i,j)$-th cell of the lattice, we fix the gauge operator $\prod_{k=0}^i Z_{k,j}Z_{k,j+1}$, whereas for a blue plaquette, we fix $\prod_{k=0}^j X_{i,k}X_{i+1,k}$. Bacon-Shor codes correspond to an empty coloring, whereas the standard surface code correspond to a red and blue checkerboard.  Note that a plaquette can be colored either red or blue, but not both, as the resulting stabilizers would not commute.

\subsection{Noise Models}

In order to carry out numerical simulations, we restrict ourselves to asymmetric Pauli noise.  We consider the \emph{$\eta$-biased depolarizing channel with error rate $p$}, defined as $$\mathcal{E}(\rho) = (1 - p)\rho + p_X X\rho X + p_Y Y \rho Y + p_Z Z \rho Z,$$ where $p = p_X + p_Y + p_Z$ and $\eta = p_Z / (p_X + p_Y)$.  We make the simplifying assumption that $p_X = p_Y$, matching the definition in \cite{Tuckett:2018}.  The notion of a physical error rate $p$ is then well-defined, as the fidelity of such a channel to the identity is independent of $\eta$.  

We consider both symmetric but biased noise, in which each qubit experiences the same error channel, as well as spatially inhomogeneous noise models, in which the error channel may depend on the qubit's position in the lattice.  In the latter case, we define the error-rate and bias of the channel on the lattice as a whole as the average fidelity and bias over each qubit in the lattice.  Note that we must be careful in comparing such models.  For example, concentrating noise on a small subset of qubits might always produce perfectly correctable errors, whereas distributing that noise symmetrically will not.

Finally, depending on the context, we consider either the \emph{code capacity} setting, in which syndrome measurements are assumed perfect, or the \emph{phenomenological setting}, in which the syndrome measurements can be faulty.
\subsection{Decoders}

Inherent to any discussion on thresholds is a choice of decoder.  We focus on three decoders: randomized, minimum-weight perfect matching, and union-find decoding. We choose among these according to our computational needs.

Each of these decoders corrects $X$- and $Z$- type errors independently.  Thus, any gains in threshold scaling are a product of the tailored syndrome information alone; it is these gains we aim to quantify.  For example, we expect that using a correlated decoder with $X$- and $Y$-type stabilizers would augment the threshold further \cite{Tuckett:2018}. 

\subsubsection{Decoder Graph}
For independent correction of $X$- and $Z$-type errors on a CSS code, the relevant decoding information is captured in the \emph{decoder (hyper-)graph}.  The decoder graph for phase errors is constructed by associating a vertex to each $X$-type stabilizer and a (hyper-)edge to each qubit, where the edges connects all stabilizers incident to that qubit. The decoder graph for bit-flip errors is defined analogously.  Note that for the subspace codes we consider, the decoder graph corresponds to a cellulation realizing that code as a homological surface code.  For an example of the phase-error decoder graph of a compass code, see Figure \ref{cut_code_graph}.

The task of a decoder is then, given some syndrome information in the form of marked nodes, identify the corresponding edge configuration producing those marked nodes \emph{up to homology}.  The question we consider in this paper is,
\\

\emph{What threshold gains do we obtain by modifying the phase and bit-flip decoder graphs according to asymmetrically distributed edge failure probabilities?}

\subsubsection{Maximum-Likelihood and Randomized Decoding}

The \emph{maximum-likelihood decoder} is one that takes in syndrome information, and chooses the most likely \emph{error-class} producing that syndrome.  Formally, its probability of success is given by $p_{succ} = \sum_{s} \Pr(\overline{E}_s),$ where the sum runs over all syndromes $s$ and $\overline{E}_s$ is the most likely error-class conditioned on syndrome $s$.  This decoder will yield optimal thresholds, but is often inefficient to implement.

The related \emph{randomized decoder} is a probabilistic decoder that, conditioned on syndrome $s$, chooses to correct error-class $\overline{E}$ with probability $\Pr(\overline{E}|s)$.  Thus, its probability of success is given by $p_{succ} = \sum_E \Pr(E) \Pr(\overline{E}|s_E)$, where $s_E$ is the syndrome associated to $E$.  The randomized decoder can well-approximate the maximum-likelihood decoder in the limit of large lattices.  Its threshold can also often be estimated thanks to a well-established connection to statistical mechanics \cite{kubica2018three, Dennis:2002}.

\subsubsection{Minimum-Weight Perfect Matching Decoding}

The minimum-weight perfect matching (MWPM) decoder assigns to each syndrome the error-class corresponding to a most-likely \emph{individual error} producing that syndrome.  Its probability of success is then $p_{succ} = \sum_s \Pr(\overline{E_s})$ where $E_s$ is a most-likely error producing syndrome $s$.

This decoder is implemented by constructing a minimum-weight perfect matching among the marked vertices in the decoder graph.  The edge weights between two marked vertices correspond to the most probable path between them; for symmetric noise, this is simply the shortest-path, but for asymmetric noise it need not be.  Fortunately, Edmond's blossom algorithm runs efficiently on graphs without hyper-edges, taking $O(n^3)$ time on a graph with $n$ nodes \cite{edmonds2009paths}.

Within the subfamily of compass codes we focus on, each qubit participates in at most two stabilizer generators.  As a result, the corresponding decoder graphs contain no hyper-edges, and so compass codes inherit the efficient MWPM decoder of the surface code.

When dealing with boundary conditions, some care must be taken to ensure a perfect-matching exists, since the parity of the marked nodes may no longer be even.  We use the techniques of \cite{Wang:2009} to estimate the logical error rates in the presence of boundaries.

\subsubsection{Asymmetrically-Weighted Union-Find Decoding}

The final decoder we will use is an asymmetrically-weighted variant of the union-find decoder recently proposed in \cite{delfosse2017almost}.  This decoder is guaranteed to performed optimally on errors of weight at most $\lfloor d/2 \rfloor$, and has been shown empirically to perform almost as well as MWPM on toric codes with respect to its threshold.  

For simplicity, our simulations are run with a periodic north-south boundary condition, which suffices for threshold comparison \cite{fowler2013accurate}.  However for completeness, we summarize the decoder on lattices with boundary, along with our modifications to account for asymmetric error rates. Decoding proceeds in two steps. 

\paragraph*{(1) Asymmetrically-Weighted Syndrome Validation.}

 The first step is (weighted) syndrome validation.  In this step, we form an erasure that is consistent with the observed syndrome \emph{and} which accounts for the asymmetric error rates.  To satisfy the first property, we save each node as a cluster, growing all clusters with an odd number of marked nodes by half-edges. After each growth, we fuse those clusters that intersect.  The cluster growth terminates when each cluster has an even number of marked nodes, indicating that we can form a hypothetical erasure that is consistent with the observed syndrome.  Furthermore, we use the weighted-growth heuristic, growing only those odd clusters in each step whose boundary is smallest.  We refer to the reader to \cite{delfosse2017almost} for a more lengthy description of syndrome-validation.

In the case of a decoder graph with boundary, we no longer have a guarantee that there are an even number of syndromes in our graph.  This is because some of the syndromes might condense at the boundary.  To accommodate for this, we simply treat the boundary as a sink in which every cluster that fuses with the boundary is assigned an even parity.  

After this, syndrome verification concludes by choosing a spanning forest within the clusters.  We asymmetrically weight syndrome verification by using Kruskal's algorithm to choose a maximum-weight spanning forest, where each edge is weighted according to its probability of failure \cite{kruskal1956shortest}.  This increases the probability of identifying the erroneous qubits.

\paragraph*{(2) Peeling With Boundaries.}
Having associated to the graph an erasure forest that is consistent with the observed syndrome and asymmetric error rates, the second step is to apply maximum-likelihood erasure decoding in the form of an altered peeling decoder \cite{delfosse2017linear}.

To each leaf node of the resulting erasure forest, we apply the rules:

\begin{itemize}
\item[($i$)] If the leaf node is marked, apply a phase flip to the corresponding edge and flip the mark of the connected node.  Then remove the leaf node and edge from the erasure tree.
\item[($ii$)] If the leaf node is unmarked, remove it and the corresponding edge from the erasure tree.
\end{itemize}

At this stage, we have an erasure forest with no leaf nodes and potentially some open edges connecting to the open boundaries.  Unfortunately, these open edges are missing their leaf nodes, and so we cannot peel.  In \cite{delfosse2017linear}, this is avoided by growing the spanning forest so that each tree has at most one open edge, and then peeling towards that edge.  However, for asymmetrically-distributed noise, a maximum-weight spanning forest might not take this form.

Instead, we can use dynamic programming to find a maximum-probability error configuration consistent with syndrome information in linear time.  Fix any tree inside the forest, with edges weighted according to their error probabilites, and root the tree at any node.  Each node in this tree  corresponds to a stabilizer, which will be either marked or unmarked.  Our aim is to identify a subset $S$ of edges that is both consistent with the syndrome information, and  has maximal failure probability.  We will then apply our phase-error correction to this set.

We proceed recursively.  To each node $v$, we will associate two values. First, we compute the maximum weight of $S_v$ for the subtree rooted at $v$ over all sub-spanning trees that include the parent edge.  Second, we compute the same maximum weight of $S_v$ over all sub-spanning trees that do not include the parent edge.  Each of these updates takes constant time, assuming that the children were previously evaluated and that $v$ has bounded degree.  Iterating over all vertices in the tree and trees within the forest, this terminates in linear time, and can be used to produce the desired $S$.

By using a tree structure, the union-find growth algorithm takes $O(n \cdot \alpha(n))$ time, where $\alpha$ is the exceptionally slow-growing inverse Ackermann's function.  However, because we find a maximum-weight spanning tree, this variant requires $O(n\log(n))$-time preprocessing.  The union-find decoder is the most time-efficient of the three decoders we consider.  

\begin{figure}[htb!]
\captionsetup[subfigure]{labelformat=empty}

        \begin{subfigure}[b]{0.24\textwidth}
                \centering
                \includegraphics[width=0.95\linewidth]{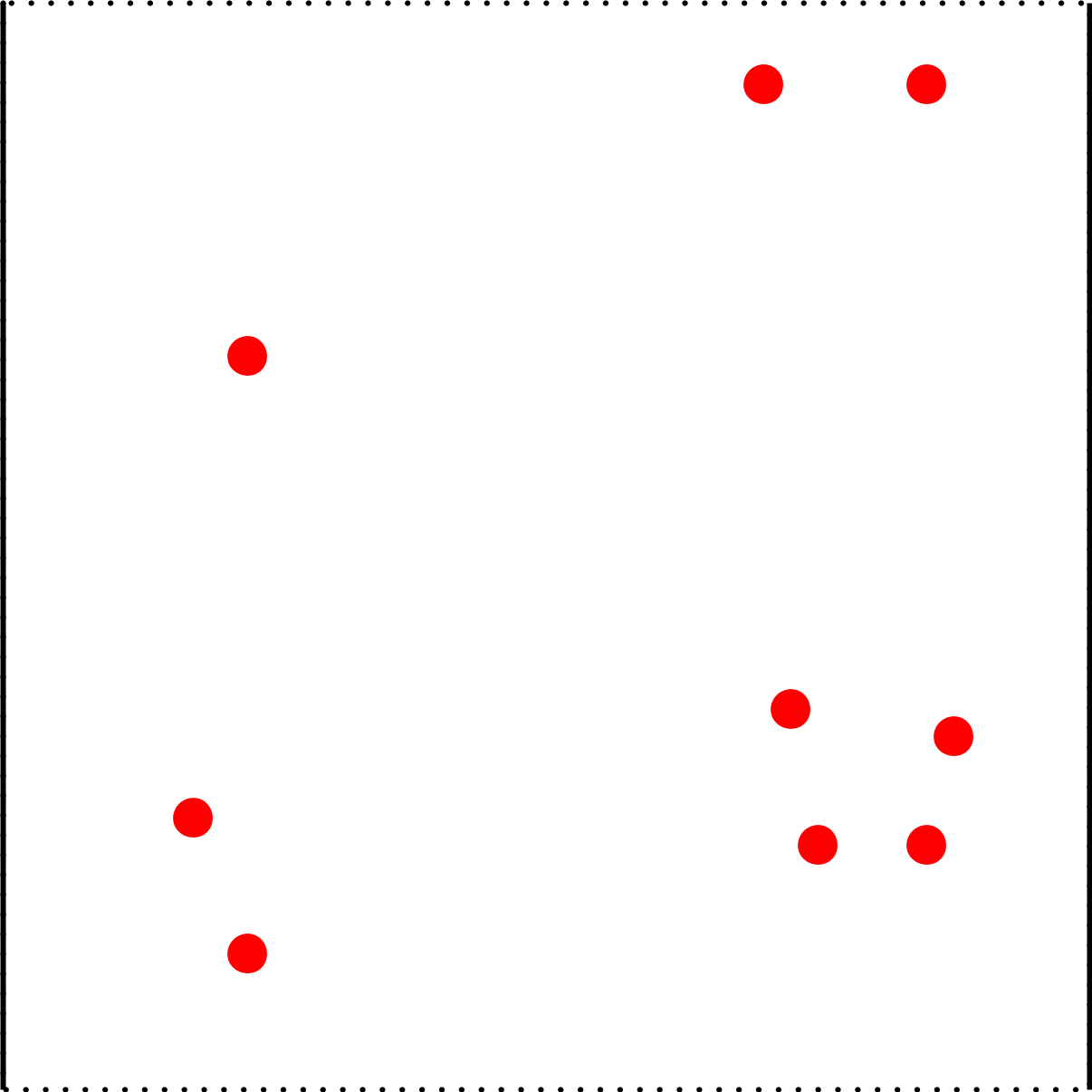}
        \end{subfigure}%
        \begin{subfigure}[b]{0.24\textwidth}
                \centering
                \includegraphics[width=0.95\linewidth]{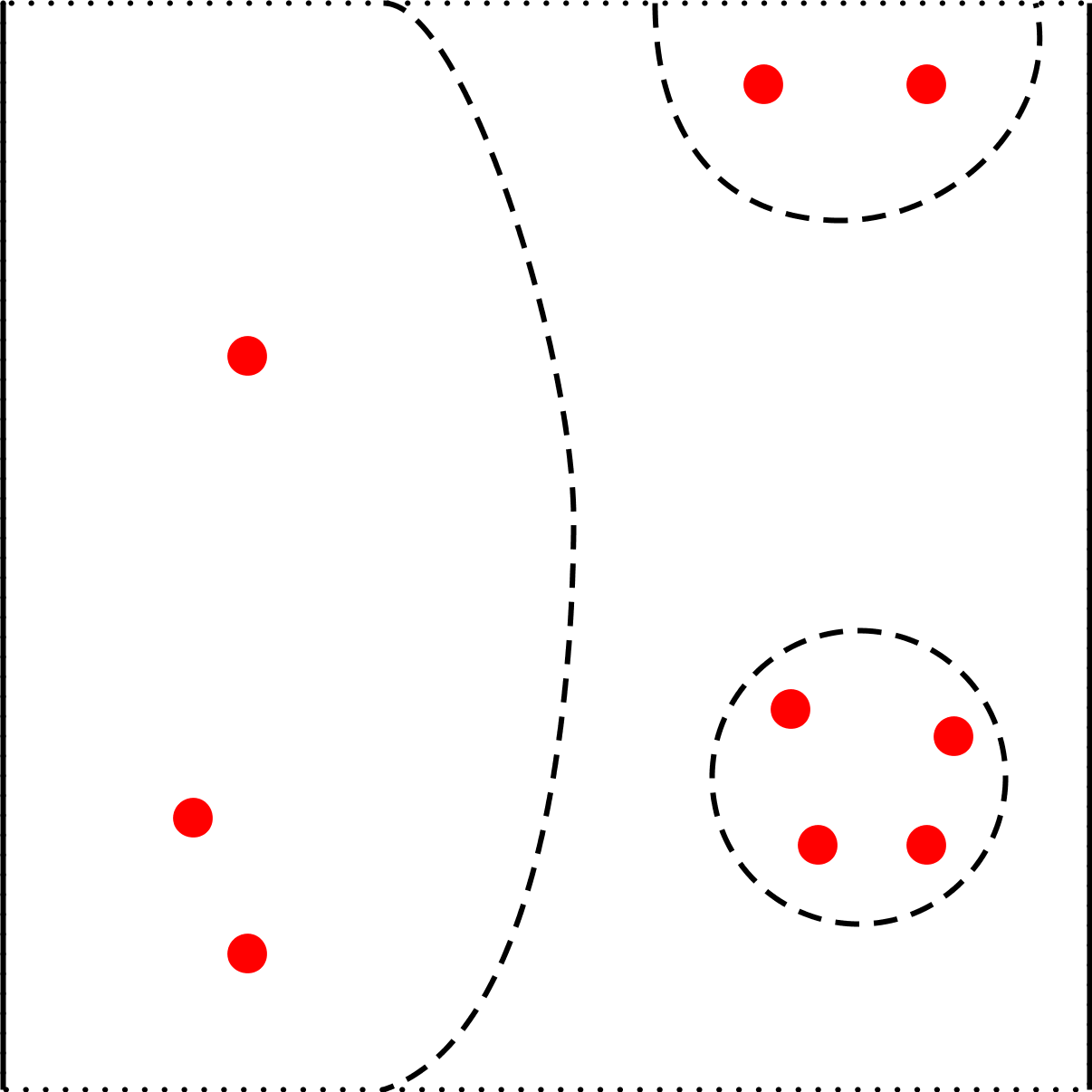}
        \end{subfigure} \par\bigskip
\begin{subfigure}[b]{0.24\textwidth}
                \centering
                \includegraphics[width=0.95\linewidth]{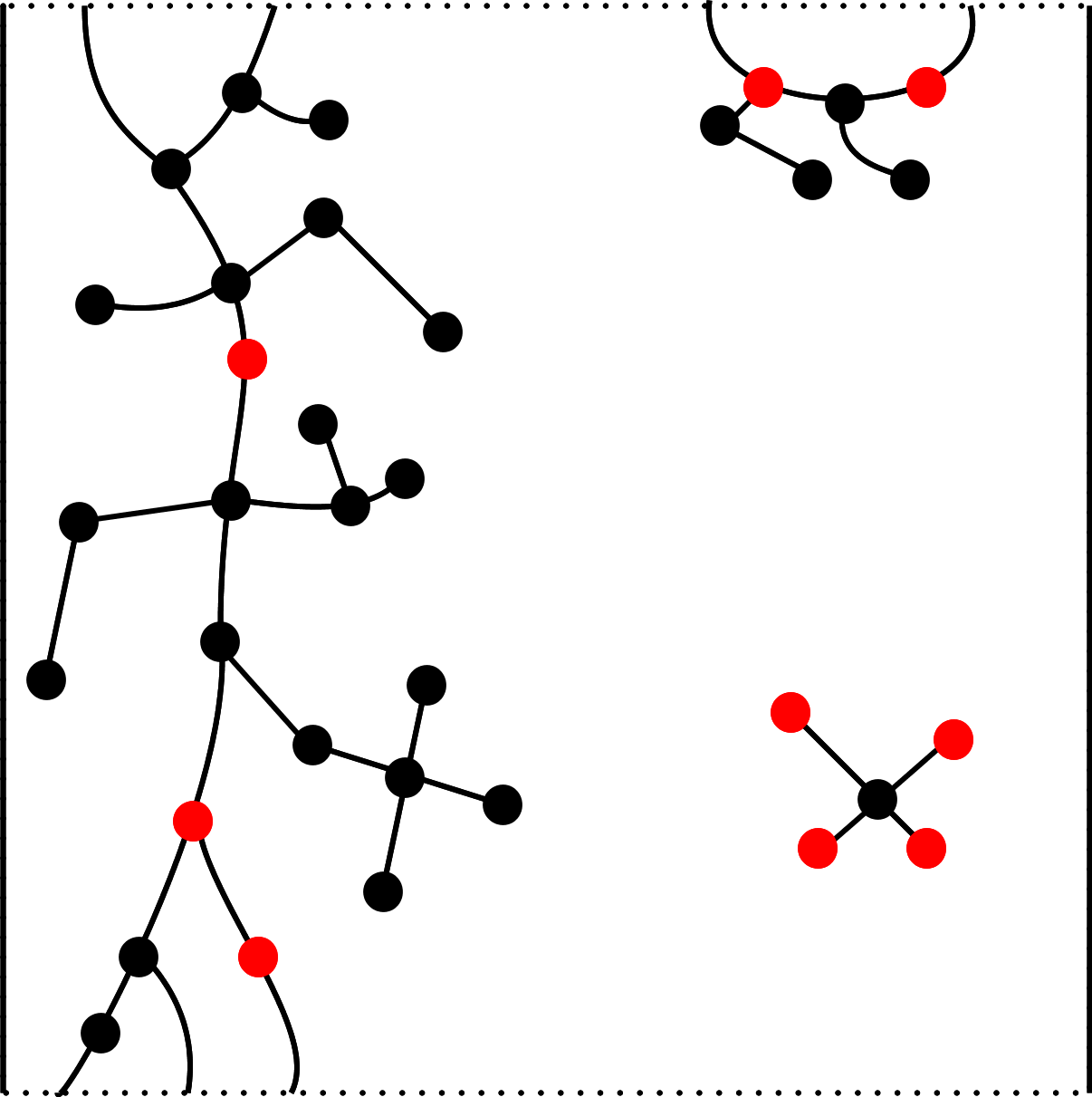}
        \end{subfigure}%
        \begin{subfigure}[b]{0.24\textwidth}
                \centering
                \includegraphics[width=0.95\linewidth]{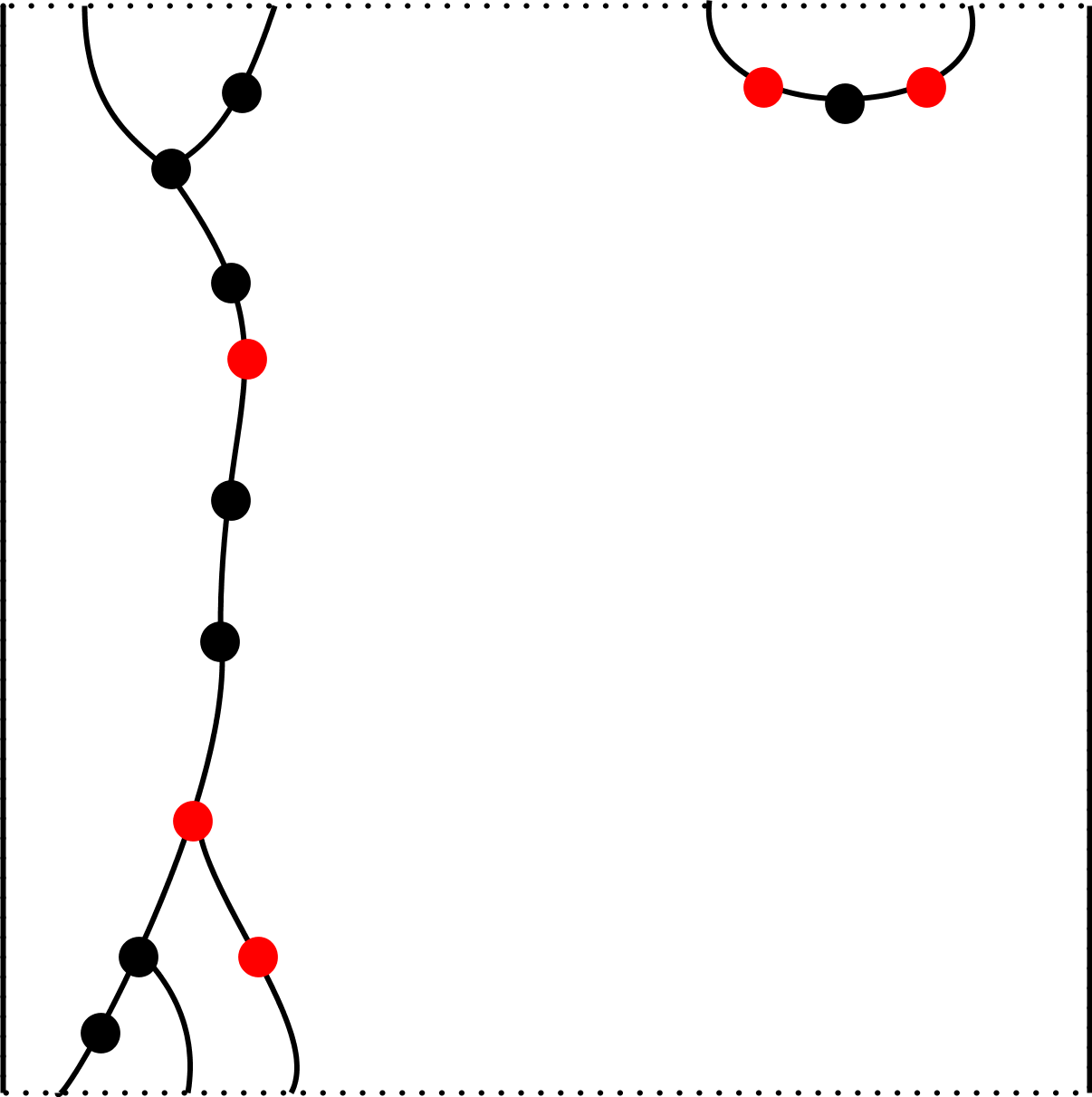}
        \end{subfigure}
        
        \caption{The square represents the decoder graph (unseen) with north-south open boundary conditions. First, the clusters (dashed enclosures) are grown to an erasure consistent with the (red) marked syndromes.  We then find maximum-weight spanning trees, with unmarked syndromes in black.  After peeling the leaf nodes, we decode the trees using dynamic programming.}
        \label{UF_Decoder}
\end{figure}

\begin{figure}[htb!]
\includegraphics[width=\linewidth]{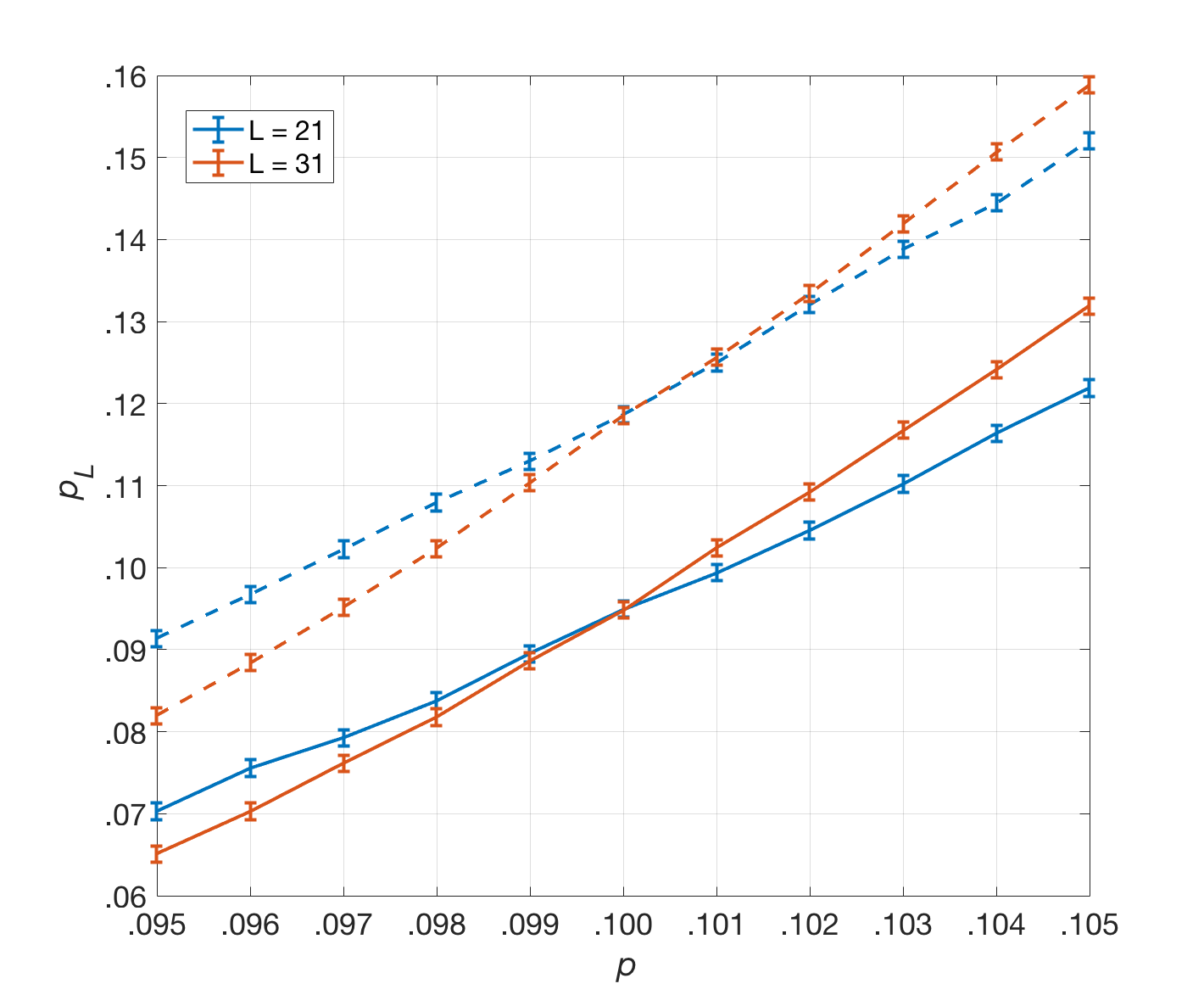}
\caption{A comparison of logical error-rates for the asymmetric decoder (solid lines) versus the symmetric decoder (dashed lines).  Each data point was generated with $10^6$ independent Monte-Carlo trials.}
\label{kruskal}
\end{figure}

For a pictoral skeleton of the decoder, see Figure \ref{UF_Decoder}.  A comparison of the decoder error-rates with and without the asymmetric alteration on the surface code is shown in Figure \ref{kruskal}.  There, the error model is generated by choosing a error probability $p_i \in_r [0,2p]$ for each physical qubit $i$ uniformly at random.  The value $p_i$ is passed to the asymmetric decoder to inform Kruskal's algorithm.  This additional information results in an improvement on the error-rate, but with little effect on the threshold.

\section{Threshold Scaling} \label{Threshold Scaling}

Before we consider asymmetric noise models, we ask the more fundamental question, \emph{how does the threshold behave in these compass codes?}  In particular, Bacon-Shor codes have no threshold while surface codes boast some of the highest thresholds.  Compass codes provide a framework for interpolating between these two, and so we examine the threshold scaling here first.

We use the code's CSS structure to argue directly about phase-flip errors of probability $p$; bit-flip errors can be decoded analogously and independently.  To correct phase errors, the relevant information about the code consists of $\mathcal{G}_Z$ and $\mathcal{S}_X$, the $Z$-type gauge subgroup and the $X$-type stabilizer subgroup, respectively.

\subsection{Surface-Density Codes}
The first family of codes we consider are the (randomized) \emph{surface-density codes}, which interpolate between the Bacon-Shor and surface code.  Each code is determined stochastically according to a \emph{surface-density} $q_{\text{surf}}$ in the following way. Given a square lattice, for each plaquette of one color in the checkerboard configuration of the surface code, we cut the corresponding $X$-type stabilizer at that plaquette with probability $q_{\text{surf}}$.  Correspondingly, $q_{\text{surf}} = 0$ is equivalent to the Bacon-Shor family (with respect to phase errors) and $q_{\text{surf}}=1$ is equivalent to the surface code.

\subsubsection{Ising Models Associated to Quantum Codes}

We identify the scaling of the threshold with the surface-density under a randomized decoder.  To do so, we exploit a connection between the threshold of quantum codes and critical temperatures of associated Ising models \cite{kubica2018three, Dennis:2002, Bombin:2010b}.

We summarize this connection briefly.  Let $\mathcal{G}_0$ be a minimal generating set of $\mathcal{G}_Z$.  Let the $g_i \in \mathcal{G}_0$ be indexed by $i$, and associate to each generator an Ising spin $s_i = \pm1$.  Index the physical qubits by $j \in \{1, \ldots, L^2\}$ and define $$ g_i(j):=
\begin{cases}
1 \hspace{0.2 cm} \text{   if $g_i$ is supported on site $j$}\\
0 \hspace{0.2 cm} \text{   otherwise.}
\end{cases}
$$
Then for any vector $\tau \in \{+1,-1\}^{L^2}$, we define the classical spin Hamiltonian
\[ H_\tau(s) = -\sum\limits_{j=1}^{L^2}\tau_j\prod\limits_{i=1}^{|\mathcal{G}_0|} s_i^{g_i(j)}. \]
For any Pauli $Z$-error $E$, define $(\tau_E)_k$ to be $-1$ if $E$ is supported on site $k$, and $+1$ otherwise.  For physical error-rate $p$, we can define the virtual temperature $\beta_p$ according to the \emph{Nishimori line} \cite{nishimori1986geometry} so that
\[\beta_p := \frac{\log(1-p) - \log(p)}{2}.\]
Define $\tau$ to be a quenched vector-valued random variable that takes value $\tau_E$ with probability $p^{|E|}(1-p)^{L^2 - |E|}$.  Under this randomly-disordered statistical model, we can express our success probability using the randomized decoder as
$$ \left\langle \left(1 + \exp{-\beta_p\cdot(F(\beta_p,\tau_{E\cdot Z_L}) - F(\beta_p, \tau_E))}\right)^{-1}\right\rangle_p$$
  where $\langle\cdot\rangle_p$ denotes the average over the random variable $\tau$ distributed according to $p$, $F$ is the free energy, and $Z_L$ is a $Z$-type representative of the logical-$Z$ operator.  In particular, finding a phase transition of the associated model at $(p_c,\beta_{p_c})$ indicates an accuracy threshold at $p_c$. 
  
\begin{figure}[htb!]
\captionsetup[subfigure]{labelformat=empty}
        \begin{subfigure}[b]{0.24\textwidth}
                \centering
                \includegraphics[width=0.95\linewidth]{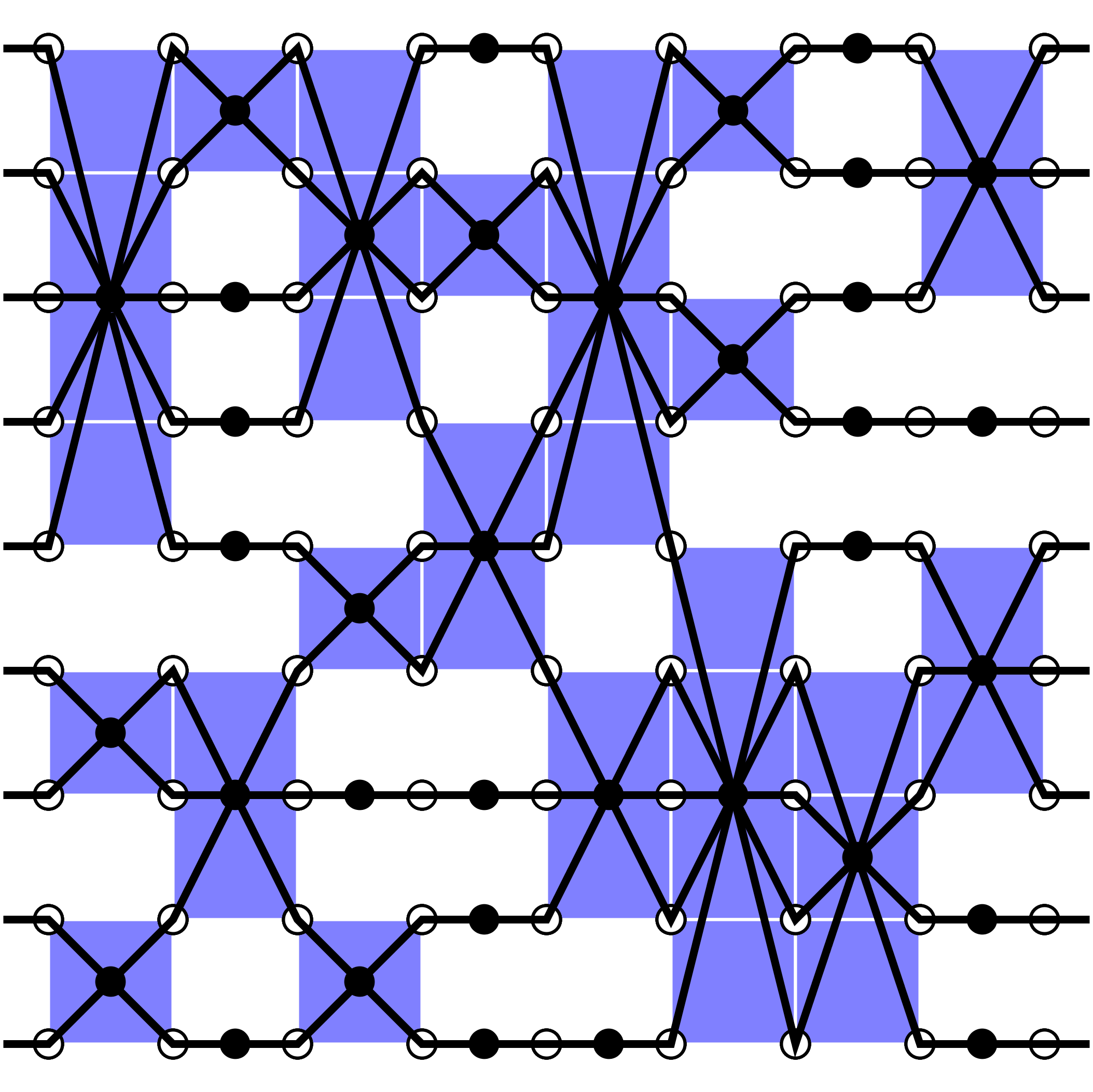}
        \end{subfigure}%
        \begin{subfigure}[b]{0.24\textwidth}
                \centering
                \includegraphics[width=0.95\linewidth]{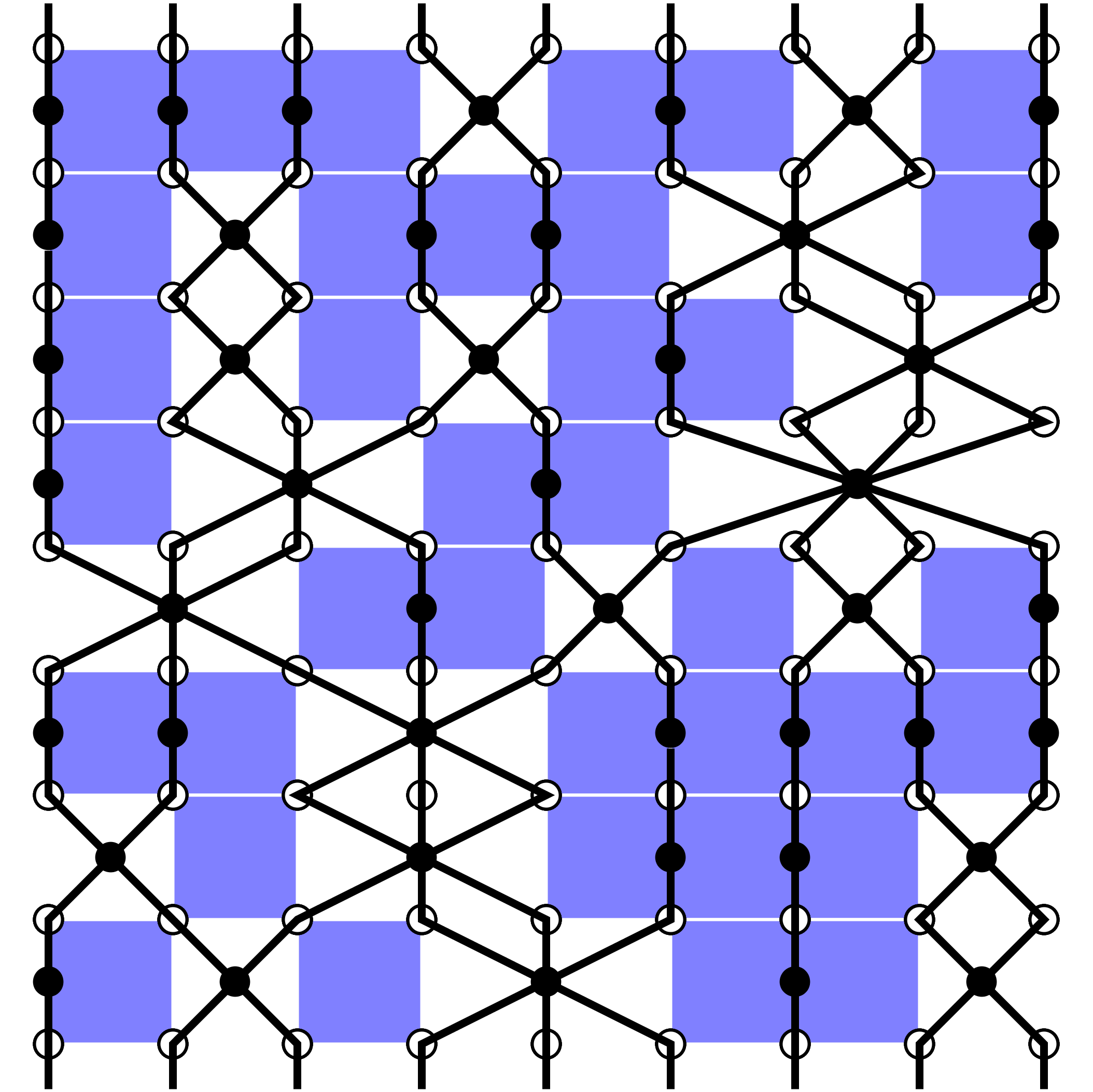}
        \end{subfigure}
        \caption{The left-hand side represents the graph describing the two-body Ising model. The right-hand side represents its dual, the decoder graph.  The blue squares represent cuts in the $X$-type stabilizers on a $9 \times 9$ lattice.  The connectivity on the left-hand graph determines the sparsity on the right.}
        \label{cut_code_graph}
\end{figure}

For an example of an Ising model associated to a compass code, see Figure \ref{cut_code_graph}.  Note that, for decoder graphs without hyper-edges, the graph defining the Ising model is dual to the decoder graph.

\subsubsection{Numerical Simulations}

\paragraph*{Parameters of the Simulation.}
We map surface-density codes to their corresponding anisotropic Ising models on random graphs. We generate random samples of the model with the given $q_{\text{surf}}$ and $p$ for various system sizes $L$, with the temperature determined by the Nishimori line according to the disorder parameter $p$. For each random trial, we use a cluster algorithm \cite{cluster} and improved estimator to compute the Binder cumulant \cite{Binder1981}. Finally, we scan over $p$ (at a separation of $0.1$ for $\ln(p)$) and look for a transition point. The system size we use ranges from $L = 5$ to $L = 61$, the number of steps for the cluster update ranges from $10^6$ to $10^8$, and the number of random trials for each parameter set ranges from $200$ to $10^4$.  

In general, as the transition point $p_c$ increases with $q_{\text{surf}}$, it enhances the frustration in the system and so more steps are needed for convergence. This is verified by the autocorrelation of the observables. However, for larger $q_{\text{surf}}$, the slope of the Binder cumulant $U$ with respect to $-\ln(p)$ also increases.  As a result, less samples and smaller system sizes are required to achieve the same level of accuracy.

\begin{figure}[htb!]
\includegraphics[width=\linewidth]{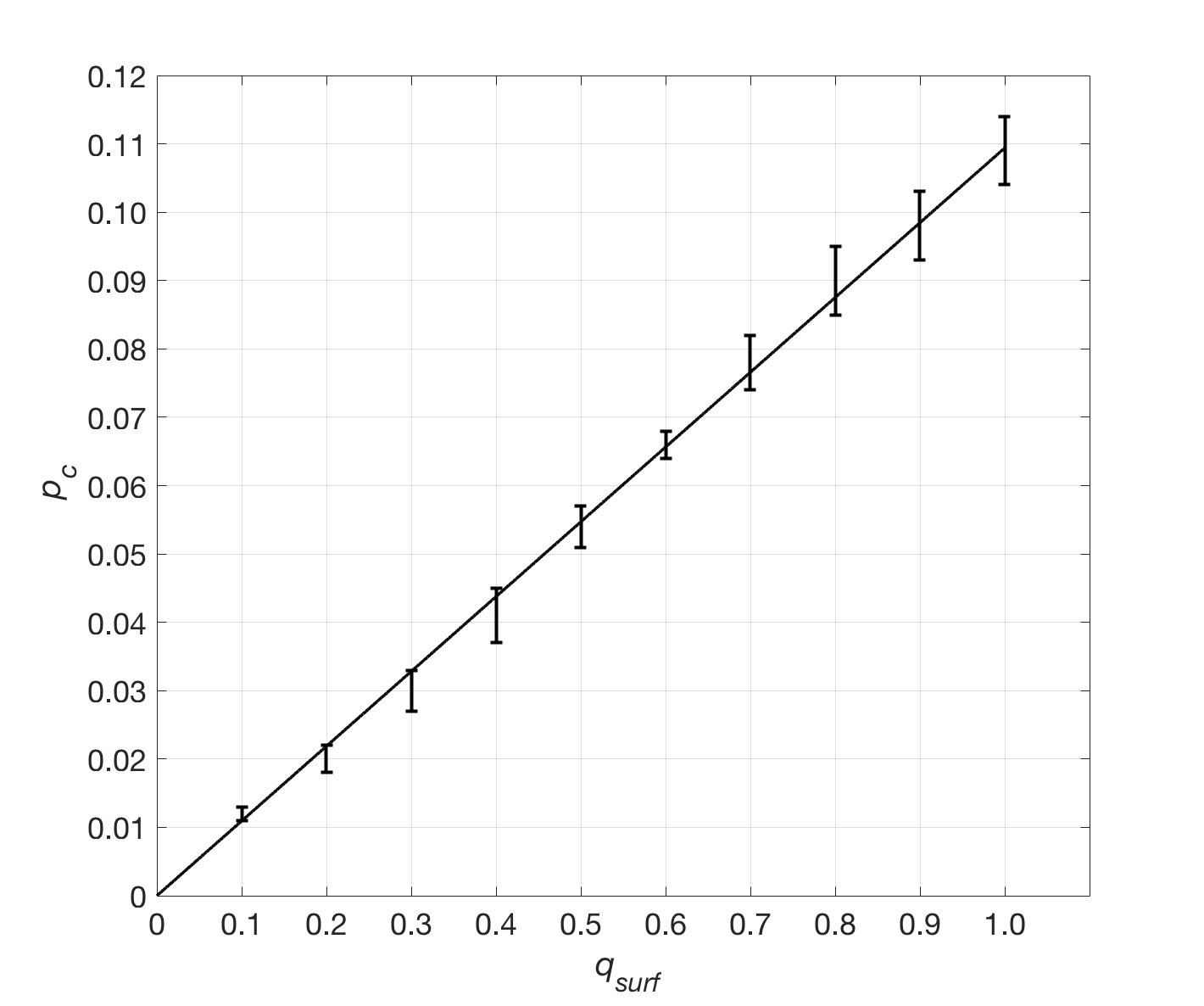}
\caption{Scaling of the critical disorder $p_c$ with respect to the surface-density $q_{\text{surf}}$.  Autocorrelation is checked using a binning analysis; fit is linear through the origin.  The widest error bars are of total width $\approx 1\%$.  At $q_{\text{surf}}=1$, the results closely match the established critical point at $p_c = 0.1094 \pm 0.0002$ \cite{Honecker:2000}.}
\label{surface_density_plot}
\end{figure}

\paragraph*{Numerical Results.}
Interestingly, simulations suggest that the threshold grows linearly with the surface density, see Figure \ref{surface_density_plot}.  In particular, a positive density is both necessary and sufficient for the presence of a threshold.  The linearity contrasts with the the threshold scaling of the less restricted code family that we consider next.

\subsection{Shor-Density Codes}

We next turn our attention to \emph{Shor-density codes}, which form a randomized family of codes that interpolate between Bacon-Shor codes and their full $X$-type gauge-fix, Shor's subspace code.  These codes are defined similarly to surface-density codes according to a new parameter which we call the \emph{Shor-density} $q_{\text{shor}}$.  For these codes, $X$-type stabilizers are cut at \emph{each} plaquette with probability $q_{\text{shor}}$.  Thus, $q_{\text{shor}}=0$ again corresponds to the Bacon-Shor code whereas $q_{\text{shor}}=1$ corresponds to Shor's subspace code \cite{Shor:1995b}.

Of course, the thresholds for such codes are one-sided: more cuts for one type of stabilizer leaves less for the other.  Consequently, such codes are best suited for asymmetric noise models.  Note that these codes remain local, in the sense that the expected maximum stabilizer weight grows logarithmically in the lattice size for any fixed $q_{\text{shor}}$. 

Because the associated graphs to these codes have a richer structure which may hinder convergence of the clustering algorithm, we instead study these codes using the union-find decoder.  We generate a new decoder graph and error in each round, and perform $10^6$ Monte Carlo trials for each data point.  We then exploit the efficiency of the union-find decoder to run $300$ Monte-Carlo trials on a $1001 \times 1001$ lattice to verify the thresholds, which should sharply converge to either $p_L = 0$ or $p_L = 0.5$ about the threshold.  This large lattice size is necessary to mitigate the growing finite-size effects.

\begin{figure}[htb!]
\includegraphics[width=\linewidth]{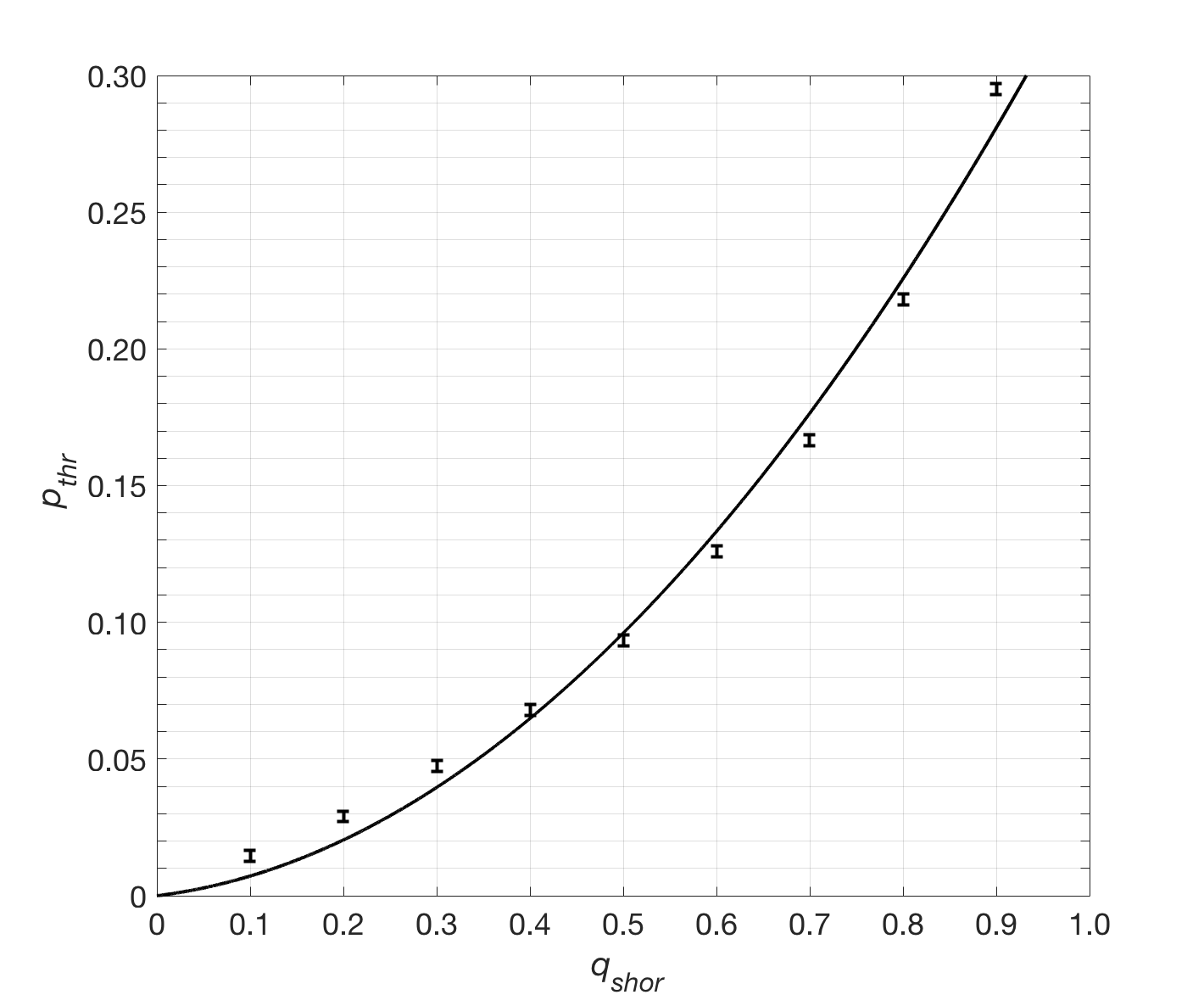}
\caption{Scaling of the estimated threshold $p$ with respect to the Shor-density $q_{\text{shor}}$.  The fit is quadratic through the origin; the finite-size effects are apparent.  All points were obtained on $81 \times 81$ size lattices, except for $q_{\text{shor}} = 0.9$.  To emphasize finite-size effects, this was performed on a $631 \times 631$ lattice, which is greater than necessary for fault-tolerant computation \cite{Fowler:2012}.}
\label{density_scaling}
\end{figure}

The threshold scaling in Figure \ref{density_scaling}, nearly saturates the zero-rate quantum Gilbert-Varshamov bound \cite{calderbank1996good}, $$H(p_x) + H(p_z) \leq 1,$$ mirroring results obtained on other lattice configurations \cite{fujii2012error, rothlisberger2012incoherent}. One thing to note is the normalizing finite-size effects at very high and very low densities.  Note that, at $q_{\text{shor}} = 1$, we essentially have disjoint copies of a repetition code. This has a threshold of $50\%$, since the union-find decoder behaves optimally on errors of weight $< \lfloor L/2 \rfloor$ \cite{delfosse2017almost}.  However, we observe a pseudothreshold of $\approx 45\%$ for union-find decoding on a $1001 \times 1001$ lattice, matching the analytical solution $$ p_{\text{logical}} = \frac{1}{2}(1 - (1-2p_\text{rep})^L)$$ where $p_\text{rep}$ is the probability of failure of a repetition code of length $L$, $$p_\text{rep} = \sum\limits_{k=\lceil\frac{L}{2}\rceil}^L \binom{L}{k}p^k(1-p)^{L-k}.$$

\paragraph*{Summary.}
These simulations suggest that the threshold is determined predominantly by the density of syndrome measurements, rather than their specific configuration, for symmetrically distributed noise.  The usual surface code does not far outperform randomized codes of equal density by this metric; it does only slightly, as its symmetry will minimize the number of $\lceil L/2 \rceil$-weight errors that introduce a logical error.  This is reinforced by the observation that the threshold appears to scale linearly with surface-density, but is strictly convex with respect to Shor-density.

\section{Asymmetric Noise} \label{Asymmetric Noise}

Next, we turn our attention to asymmetric noise.  We consider two different types: biased noise that is symmetrically distributed throughout the lattice, and asymmetrically distributed noise.  In both cases, we find that substantial gains can be made by tailoring the decoder graphs to the noise directly.  We analyze these in both the code capacity and phenomenological models, and compute their thresholds under different noise biases.

\subsection{Biased but Symmetric Noise}

For biased noise that is symmetrically distributed, we construct a family of compass codes we call \emph{elongated codes}.  These codes are defined by a parameter $\ell \in \mathbb{N}^+$ we call the \emph{elongation} of the code, and are constructed by cutting the $Z$-stabilizers at the $(i,j)$-th plaquettes for all $i-j \equiv 0$ $(\text{mod }\ell)$.  The $X$-stabilizers are then cut at all remaining plaquettes, resulting in a subspace code.  This is similar to the approaches of \cite{Aliferis:2007,stephens2013high}, which consider concatenations with different phase-flip repetition codes.

Under this definition, we obtain Shor's code for $\ell = 1$ and the surface code for $\ell = 2$.  For $\ell > 2$, we obtain an asymmetrization of Kitaev's toric code in the bulk with extended $2\ell$-body plaquette operators.  This family illustrates a simple compass code is well-equipped to correct asymmetric noise, while sacrificing somewhat in locality. 

It is worth noting that choosing asymmetric lattice dimensions as in \cite{Napp:2012} may alter the logical error-rate of a code family, but will not change the threshold, as it is a property of the bulk.  Thus, the elongation of the code refers to a stretching of the bulk stabilizer geometry, not the lattice itself.

As the elongation grows, finite-size effects play a greater role.  As such, we use MWPM decoding to perform simulations on smaller lattices at lower elongations, and union-find decoding to test larger lattices.  While these larger lattices also suffer from finite-size effects, we use the efficiency of union-find decoding to simulate lattices of between $10^3$ and $10^4$ qubits, which is the estimated code-size required for full-scale fault-tolerant computation \cite{Fowler:2012}.

Furthermore, we estimate the phenomenological threshold by simulating $(2+1)$-$D$ elongated codes.  For a physical lattice of linear size $L$, this corresponds to performing $L$ rounds of faulty syndrome extraction, followed by an ideal round, and then decoding.  The corresponding decoder graph is then $L+1$ copies of the initial decoder graph with $L$ time-like slices of edges connecting the corresponding vertices in each space-like slice.  These time-like edges represent faulty measurements.  

Although the size of each stabilizer is independent of the lattice size, we scale the probability of failure for each stabilizer linearly with its weight.  We assume the usual phenomenological normalization that plaquette stabilizers are faulty at the physical error rate $p$. Despite some increasing stabilizer weights, we observe substantial threshold gains in both the code capacity \emph{and} phenomenological models.  

Tables \ref{mwpm_cc} and \ref{uf_cc} show the code-capacity thresholds using the MWPM and union-find decoder, respectively, while Table \ref{uf_ph} shows the phenomenological threshold using the union-find decoder.  In these tables, $\eta_{\text{opt}}$ refers to the optimal bias that realizes the threshold $p_{\text{thr}}$, while $\eta_*$ is the bias above which the codes will outperform the surface code.

\bgroup 
\def\arraystretch{1.6}
\begin{table}[htb!]
 \centerline{\begin{tabular}{|c|c|c|c|c|c|}
    \hline
   $\ell$ & $\eta_\text{opt}$ & $p_\text{thr}$ & $\eta_*$ & $p_z$ & $p_x$\\
    \hline\hline
  $2$ & $0.5$ & $15.5\%$ & $N/A$ & $10.3\% \pm 0.2 \%$& $10.3\% \pm 0.2 \%$\\
  \hline
  $3$ & $1.67$ & $17.9\%$ & $1.39$ & $14.1\% \pm 0.3 \%$& $6.5\% \pm 0.2 \%$\\
  \hline
  $4$ & $3.00$ & $20.0\%$ & $2.10$ & $17.5\% \pm 0.2 \%$& $5.0\% \pm 0.2 \%$\\
  \hline
  $5$ & $4.26$ & $21.6\%$ & $2.78$ & $19.5\% \pm 0.1 \%$& $4.1\% \pm 0.1 \%$\\
  \hline
  $6$ & $5.89$ & $22.8\%$ & $3.70$ & $21.1\% \pm 0.1 \%$& $3.3\% \pm 0.1\%$\\
  \hline
  \end{tabular}}
  \caption{Thresholds for the MWPM decoder in the code-capacity model.  Simulations were done on lattices of size at most $17\times 17$.}
\label{mwpm_cc}
\end{table}
\egroup

\bgroup 
\def\arraystretch{1.6}
\begin{table}[htb!]
 \centerline{\begin{tabular}{|c|c|c|c|c|c|}
    \hline
   $\ell$ & $\eta_\text{opt}$ & $p_\text{thr}$ & $\eta_*$ & $p_z$ & $p_x$\\
    \hline\hline
  $2$ & $0.5$ & $15.0\%$ & $N/A$ & $10.0\% \pm 0.2 \%$& $10.0\% \pm 0.2 \%$\\
  \hline
  $3$ & $1.41$ & $16.9\%$ & $1.14$ & $13.4\% \pm 0.3 \%$& $7.0\% \pm 0.2 \%$\\
  \hline
  $4$ & $2.40$ & $18.4\%$ & $1.78$ & $15.7\% \pm 0.2 \%$& $5.4\% \pm 0.2 \%$\\
  \hline
  $5$ & $3.45$ & $19.6\%$ & $2.41$ & $17.4\% \pm 0.1 \%$& $4.4\% \pm 0.1 \%$\\
  \hline
  $6$ & $4.45$ & $20.7\%$ & $2.95$ & $18.8\% \pm 0.1 \%$& $3.8\% \pm 0.1\%$\\
  \hline
  $7$ & $5.62$ & $21.9\%$ & $3.55$ & $20.2\% \pm 0.1 \%$& $3.3\% \pm 0.1\%$\\
  \hline
  $8$ & $6.23$ & $22.8\%$ & $3.84$ & $21.2\% \pm 0.1 \%$& $3.1\% \pm 0.1\%$\\
  \hline
  $9$ & $7.29$ & $23.7\%$ & $4.17$ & $22.2\% \pm 0.1 \%$& $2.9\% \pm 0.1\%$\\
  \hline
  $10$ & $8.36$ & $24.0\%$ & $4.77$ & $22.7\% \pm 0.1 \%$& $2.6\% \pm 0.1\%$\\
  \hline
  $20$ & $20.7$ & $28.3\%$ & $10.5$ & $27.6\% \pm 0.1 \%$& $1.3\% \pm 0.1\%$\\
  \hline
  $50$ & $55.3$ & $33.8\%$ & $24.0$ & $33.5\% \pm 0.1 \%$& $0.6\% \pm 0.1\%$\\
  \hline
  \end{tabular}}
  \caption{Thresholds for the union-find decoder in the code-capacity model.  Simulations were done on lattices of size at most $81 \times 81$.}
\label{uf_cc}
\end{table}
\egroup

\bgroup 
\def\arraystretch{1.6}
\begin{table}[htb!]
 \centerline{\begin{tabular}{|c|c|c|c|c|c|}
    \hline
   $\ell$ & $\eta_\text{opt}$ & $p_\text{thr}$ & $\eta_*$ & $p_z$ & $p_x$\\
    \hline\hline
  $2$ & $0.5$ & $3.98\%$ & $N/A$ & $2.65\% \pm 0.2 \%$& $2.65\% \pm 0.2 \%$\\
  \hline
  $3$ & $1.20$ & $4.45\%$ & $0.99$ & $3.4\% \pm 0.2 \%$& $2.0\% \pm 0.2 \%$\\
  \hline
  $4$ & $1.88$ & $4.60\%$ & $1.49$ & $3.8\% \pm 0.2 \%$& $1.6\% \pm 0.2 \%$\\
  \hline
  $5$ & $2.73$ & $4.85\%$ & $2.06$ & $4.2\% \pm 0.2 \%$& $1.3\% \pm 0.2 \%$\\
  \hline
  $6$ & $3.17$ & $5.00\%$ & $2.32$ & $4.4\% \pm 0.2 \%$& $1.2\% \pm 0.2\%$\\
  \hline
  \end{tabular}}
  \caption{Thresholds for the union-find decoder in the phenomenological model.  Simulations were done on lattices of size at most $35 \times 35 \times 35$.}
\label{uf_ph}
\end{table}
\egroup

Notably, a relatively \emph{smaller} noise bias is required to outperform the surface code in the phenomenological setting.  Unsurprisingly, the MWPM outperforms the union-find decoder as a whole, but surprisingly, displays lower thresholds on lattices comprised of higher-weight stabilizers.  This suggests that union-find decoding may better exploit the degeneracy of certain lattices; in particular, one should use MWPM for $Z$-type errors and union-find decoding for $X$-type errors on elongated lattices.  Our estimates for established surface code thresholds match those found in \cite{Wang:2009} at $10.3 \%$ for MWPM decoding and in \cite{delfosse2017almost} at $9.95\%$ and $2.65\%$ for union-find decoding in $2$- and $(2+1)$-$D$, respectively.

\subsection{Spatially Dependent Noise} \label{Spatially Correlated Noise}

We conclude by considering noise that is asymmetrically distributed throughout the lattice.  To illustrate the idea, we focus on a simple noise model in which dephasing noise decays linearly from the right-hand side of the lattice according to the function $p_z(i,j) = (w(j/L) + (1-w)(1 - j/L))\cdot p_{\text{tot}}/2$.  Here, $i$ and $j$ are the coordinates of a qubit, $L$ is the linear size of the lattice, and $w$ is a constant that determines the degree of incline.  We further assume that $p_x = p_{\text{tot}}/2$, so that $p_{\text{tot}}$ is the total infidelity of the channel.  Note that the average bias between the dephasing noise and bit-flip noise is symmetric.

The idea is simple: when the noise is distributed asymmetrically, the stabilizer information can be chosen to match the noise.  Intuitively, lower weight stabilizers add more error-information about the qubits nearby.  With this in mind, we define a randomized family of codes we call \emph{($p_z$-)tailored codes}.  At each plaquette, we choose to cut the corresponding $X$-type stabilizer with probability $2p_z(i,j)/p_{\text{tot}}$, where $i,j$ are the coordinates of the upper-left qubit at that plaquette.  Then, in the presence of a high amount of dephasing noise, many low-weight $X$-type stabilizers will appear to aide in error-correction.  

\begin{figure}[htb!]
  \centering
  \begin{tabular}{@{}c@{}}
    \includegraphics[width=.854\linewidth]{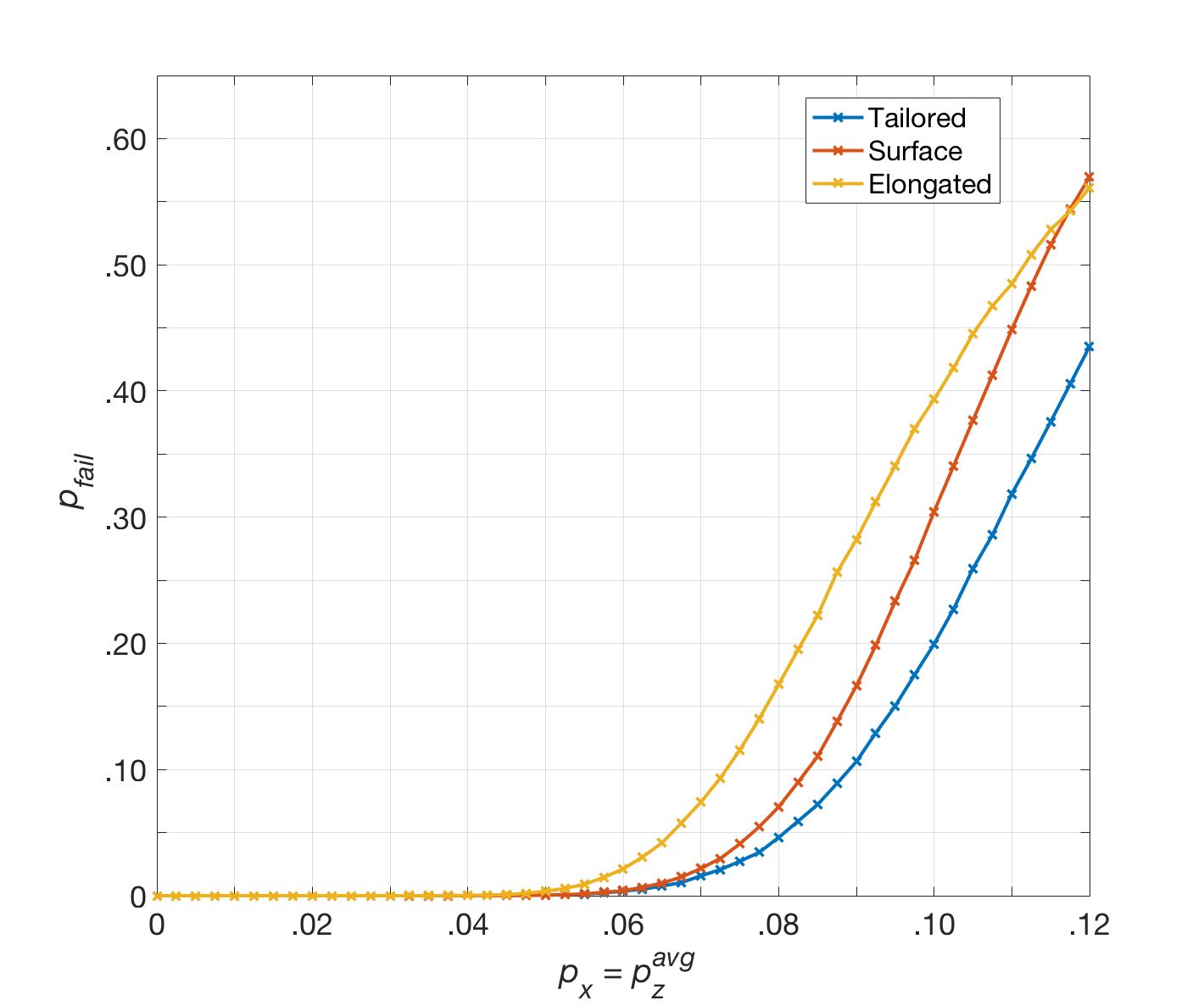}
  \end{tabular}
  
  \begin{tabular}{@{}c@{}}
    \includegraphics[width=.854\linewidth]{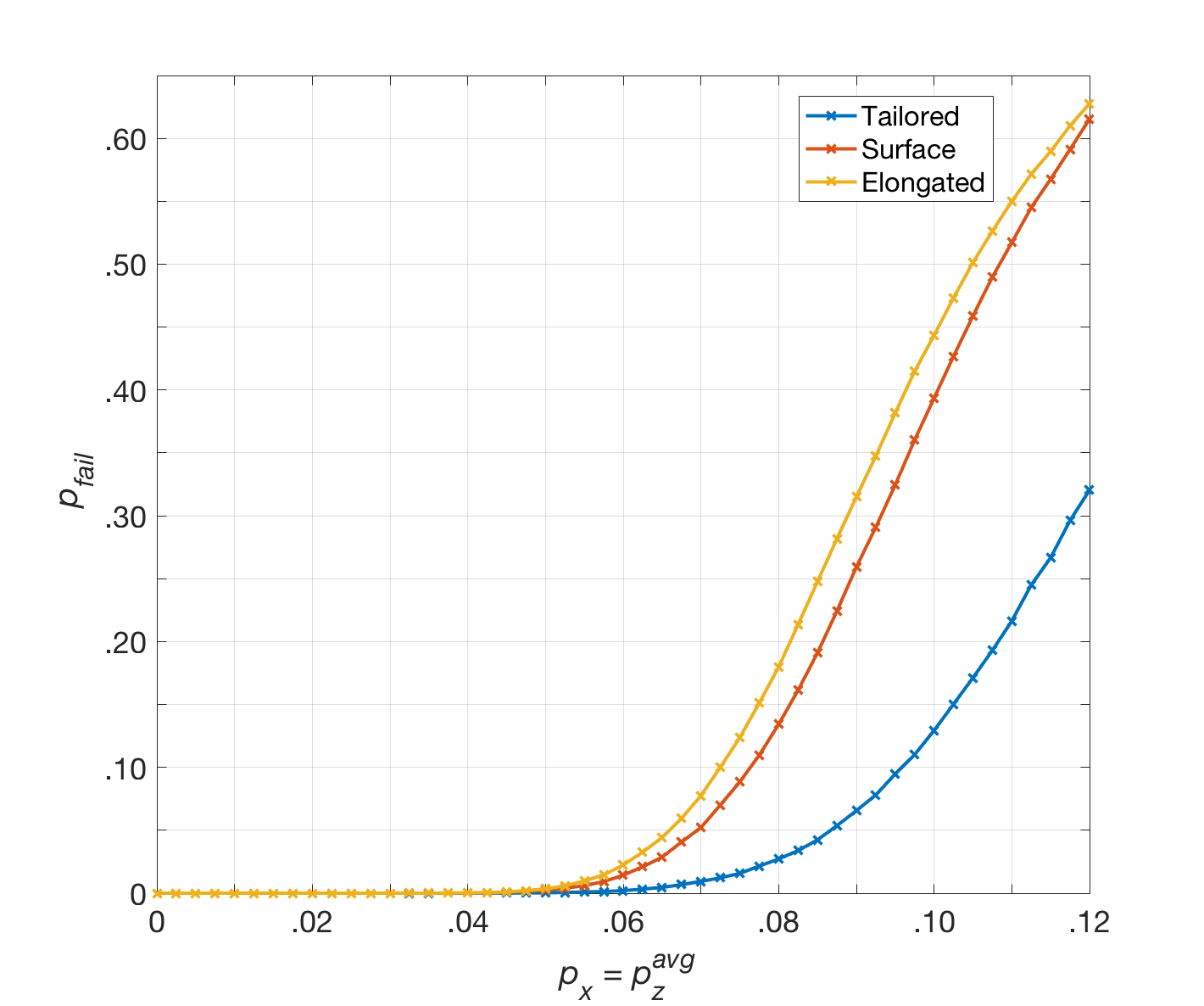}
  \end{tabular}

  \caption{Error-rates for gradually ($w = 0.25$, top) and steeply ($w = 0.10$, bottom) inclined linear noise, computed on a lattice of size $33 \times 33$ using union-find decoding in the high-noise regime.  Here, $p_{\text{fail}}$ represents the total probability of a failure in either the $X$- or $Z$-type decoders.}\label{tailored}
\end{figure}

We observe that the tailoring of these codes to the noise model can augment error rates, see Figure \ref{tailored}.  It is worth noting, however, that simply weighting the probability of each cut according to the surrounding qubits may not always be the optimal strategy.  In particular, in the low error-rate limit, this will become an optimization problem that seeks to minimize the weights of uncorrectable paths of length $\lceil \frac{L}{2} \rceil$ in the decoder graph. 

\section{Fault-tolerance with Bare Ancilla}\label{ft}
One of the major advantages that comes with the locality of the Bacon-Shor code is fault-tolerant bare-ancilla syndrome extraction \cite{Aliferis:2007, Li:2018}.  Although this extraction scheme is the simplest and least resource-intensive, most codes incur some loss in effective distance due to high-weight correlated errors produced by errors on the ancilla.  For the standard and rotated surface codes, these ``hook'' errors can be carefully designed to ensure no significant loss in performance
\cite{Tomita:2014, Dennis:2002}. 

In the compass code framework, this resilience to correlated errors is a general phenomenon resulting from measuring stabilizers along the Bacon-Shor gauge operators.  Using such a syndrome extraction scheme on any gauge-fix of the Bacon-Shor code, any collection of $d-1$ faults in the circuit produce an error of the form $EG$, where $|\text{supp}(E)| \leq d-1$ is minimal and $G$ is a gauge operator of the initial Bacon-Shor code.  

Divide the generators of the stabilizer group of any compass code into $\mathcal{S} = \langle \mathcal{S}_B,\mathcal{S}_F\rangle$, where $\mathcal{S}_B$ are the stabilizer generators of the Bacon-Shor code and $\mathcal{S}_F$ are those gauge operators that have been fixed.  Then, for any error $EG$ resulting from $d-1$ faults in the circuit, if $|\text{supp}(E)| = 0$, then either $G \in \mathcal{S}_F$ or there exists an $S \in \mathcal{S}_F: SG = -GS.$  Else if $0 < |\text{supp}(E)| < d$, then there exists an $S \in \mathcal{S}_B: SE = -ES$.  Since $S$ must also commute with any gauge operator $G$, it follows that $EG$ is detectable.  Thus, any error resulting from $\leq d-1$ faults during syndrome extraction is either detectable, or trivial.

This demonstrates that there exists fault-tolerant decoding without a loss in effective distance.  However, it is \emph{not} necessarily maximum-likelihood decoding on the memory.  One simple counter-example is Shor's code, where a single well-placed ancilla error can effect a weight $d$ memory error that maximum-likelihood will misdiagnose as a weight $d-1$ memory error, resulting in failure.  The above does imply that performing MWPM with respect to linear-probability faults in the decoder graph is fault-tolerant.  Introducing these faults amount to triangulating the decoding graph, similar to hook errors in the surface code case \cite{Wang:2009, Dennis:2002}.  Determining circuit-level compass code performance in this model is the subject of future inquiry \cite{Huang:2018}.

\section{Conclusions} \label{Discussion}

In this article, we have described an ansatz for designing planar codes stemming from the $2$-D compass model.  We have provided evidence that simple subfamilies of this class may be useful for correcting biased noise in idealized code capacity and phenomenological noise models, particularly if that bias is distributed geometrically.  In particular, one can bias the stabilizers locally towards correcting a certain error-type.

There are two central challenges for these codes in the more realistic circuit-level noise model.  Although these codes are still local, there is a trade-off between the bias of the codes and the locality of the stabilizer measurements.  We have demonstrated that fault-tolerant measurement in Bacon-Shor \cite{Aliferis:2007, Li:2018} and surface codes \cite{Dennis:2002, Tomita:2014} using bare ancilla can be adapted to the compass model, if measurements are performed in the correct order.  Nevertheless, these correlated errors will deteriorate code performance as higher-weight stabilizer outcomes become less reliable.  This might be mitigated by using other flag-type schemes, or by preserving some gauge degrees of freedom. We would expect that these gains would persist, but at the expense of higher bias and code overhead.  As such, we leave a more involved circuit-level analysis to future work.

The second concern is whether the biased noise model itself can persist at the circuit level.  To remain experimentally motivated, one must choose operations that preserve the bias \cite{PhysRevA.78.052331, Tuckett:2018,stephens2013high}.  Consequently, the construction of \emph{simple} and \emph{bias-preserving} fault-tolerant gadgets is key to utilizing asymmetric noise.

Finally, we have only narrowly broached the design space offered by these codes.  Exploring different configurations according to other geometrically-defined noise \cite{delfosse2016linear}, generalizing to codes defined on the 3-D compass model, and using correlated decoders \cite{delfosse2014decoding,nickerson2017analysing,Tuckett:2018, darmawan2018efficient, maskara2018advantages} are all avenues to explore.  More generally, finding other LDPC constructions adapted to biased noise may give the best of both worlds, mitigating the overhead of asymmetrization while taking advantage of the bias.

\section{Acknowledgements}

The authors thank Dave Bacon, Nicolas Delfosse, and Pavithran Iyer for useful discussions, and Luming Duan for providing computational resources for simulations on the Flux cluster at the University of Michigan.  Additionally, they thank anonymous referees for their helpful comments.  This research was supported in part by NSF (1717523), ODNI/IARPA LogiQ program (W911NF-10-1-0231), ARO MURI (W911NF-16-1-0349), EPiQC - an NSF Expedition in Computing (1730104), and the Alexander von Humboldt Foundation.

\bibliographystyle{apsrev}
\bibliography{bibliography}

\begin{thebibliography}{52}
\expandafter\ifx\csname natexlab\endcsname\relax\def\natexlab#1{#1}\fi
\expandafter\ifx\csname bibnamefont\endcsname\relax
  \def\bibnamefont#1{#1}\fi
\expandafter\ifx\csname bibfnamefont\endcsname\relax
  \def\bibfnamefont#1{#1}\fi
\expandafter\ifx\csname citenamefont\endcsname\relax
  \def\citenamefont#1{#1}\fi
\expandafter\ifx\csname url\endcsname\relax
  \def\url#1{\texttt{#1}}\fi
\expandafter\ifx\csname urlprefix\endcsname\relax\def\urlprefix{URL }\fi
\providecommand{\bibinfo}[2]{#2}
\providecommand{\eprint}[2][]{\url{#2}}

\bibitem[{\citenamefont{Aliferis et~al.}(2005)\citenamefont{Aliferis,
  Gottesman, and Preskill}}]{Aliferis:2006}
\bibinfo{author}{\bibfnamefont{P.}~\bibnamefont{Aliferis}},
  \bibinfo{author}{\bibfnamefont{D.}~\bibnamefont{Gottesman}},
  \bibnamefont{and} \bibinfo{author}{\bibfnamefont{J.}~\bibnamefont{Preskill}},
  \bibinfo{journal}{Quantum Inf. Comput.} \textbf{\bibinfo{volume}{6}},
  \bibinfo{pages}{97} (\bibinfo{year}{2005}).

\bibitem[{\citenamefont{Knill et~al.}(1996)\citenamefont{Knill, Laflamme, and
  Zurek}}]{Knill:1996b}
\bibinfo{author}{\bibfnamefont{E.}~\bibnamefont{Knill}},
  \bibinfo{author}{\bibfnamefont{R.}~\bibnamefont{Laflamme}}, \bibnamefont{and}
  \bibinfo{author}{\bibfnamefont{W.}~\bibnamefont{Zurek}},
  \bibinfo{journal}{arXiv preprint quant-ph/9610011}  (\bibinfo{year}{1996}).

\bibitem[{\citenamefont{Aharonov and Ben-Or}(1997)}]{Aharonov:1997}
\bibinfo{author}{\bibfnamefont{D.}~\bibnamefont{Aharonov}} \bibnamefont{and}
  \bibinfo{author}{\bibfnamefont{M.}~\bibnamefont{Ben-Or}}, in
  \emph{\bibinfo{booktitle}{Proceedings of the twenty-ninth annual ACM
  symposium on Theory of computing}} (\bibinfo{organization}{ACM},
  \bibinfo{year}{1997}), pp. \bibinfo{pages}{176--188}.

\bibitem[{\citenamefont{Dennis et~al.}(2002)\citenamefont{Dennis, Kitaev,
  Landahl, and Preskill}}]{Dennis:2002}
\bibinfo{author}{\bibfnamefont{E.}~\bibnamefont{Dennis}},
  \bibinfo{author}{\bibfnamefont{A.}~\bibnamefont{Kitaev}},
  \bibinfo{author}{\bibfnamefont{A.}~\bibnamefont{Landahl}}, \bibnamefont{and}
  \bibinfo{author}{\bibfnamefont{J.}~\bibnamefont{Preskill}},
  \bibinfo{journal}{Journal of Mathematical Physics}
  \textbf{\bibinfo{volume}{43}}, \bibinfo{pages}{4452} (\bibinfo{year}{2002}).

\bibitem[{\citenamefont{Bomb{\'\i}n}(2015)}]{Bombin:2013}
\bibinfo{author}{\bibfnamefont{H.}~\bibnamefont{Bomb{\'\i}n}},
  \bibinfo{journal}{New Journal of Physics} \textbf{\bibinfo{volume}{17}},
  \bibinfo{pages}{083002} (\bibinfo{year}{2015}).

\bibitem[{\citenamefont{Tomita and Svore}(2014)}]{Tomita:2014}
\bibinfo{author}{\bibfnamefont{Y.}~\bibnamefont{Tomita}} \bibnamefont{and}
  \bibinfo{author}{\bibfnamefont{K.~M.} \bibnamefont{Svore}},
  \bibinfo{journal}{Physical Review A} \textbf{\bibinfo{volume}{90}},
  \bibinfo{pages}{062320} (\bibinfo{year}{2014}).

\bibitem[{\citenamefont{Yoder and Kim}(2017)}]{Yoder:2017b}
\bibinfo{author}{\bibfnamefont{T.~J.} \bibnamefont{Yoder}} \bibnamefont{and}
  \bibinfo{author}{\bibfnamefont{I.~H.} \bibnamefont{Kim}},
  \bibinfo{journal}{Quantum} \textbf{\bibinfo{volume}{1}}, \bibinfo{pages}{2}
  (\bibinfo{year}{2017}).

\bibitem[{\citenamefont{Fowler et~al.}(2012)\citenamefont{Fowler, Mariantoni,
  Martinis, and Cleland}}]{Fowler:2012}
\bibinfo{author}{\bibfnamefont{A.~G.} \bibnamefont{Fowler}},
  \bibinfo{author}{\bibfnamefont{M.}~\bibnamefont{Mariantoni}},
  \bibinfo{author}{\bibfnamefont{J.~M.} \bibnamefont{Martinis}},
  \bibnamefont{and} \bibinfo{author}{\bibfnamefont{A.~N.}
  \bibnamefont{Cleland}}, \bibinfo{journal}{Physical Review A}
  \textbf{\bibinfo{volume}{86}}, \bibinfo{pages}{032324}
  (\bibinfo{year}{2012}).

\bibitem[{\citenamefont{Terhal}(2015)}]{terhal2015quantum}
\bibinfo{author}{\bibfnamefont{B.~M.} \bibnamefont{Terhal}},
  \bibinfo{journal}{Reviews of Modern Physics} \textbf{\bibinfo{volume}{87}},
  \bibinfo{pages}{307} (\bibinfo{year}{2015}).

\bibitem[{\citenamefont{O’brien et~al.}(2017)\citenamefont{O’brien,
  Tarasinski, and DiCarlo}}]{o2017density}
\bibinfo{author}{\bibfnamefont{T.}~\bibnamefont{O’brien}},
  \bibinfo{author}{\bibfnamefont{B.}~\bibnamefont{Tarasinski}},
  \bibnamefont{and} \bibinfo{author}{\bibfnamefont{L.}~\bibnamefont{DiCarlo}},
  \bibinfo{journal}{npj Quantum Information} \textbf{\bibinfo{volume}{3}},
  \bibinfo{pages}{39} (\bibinfo{year}{2017}).

\bibitem[{\citenamefont{Trout et~al.}(2018)\citenamefont{Trout, Li,
  Guti{\'e}rrez, Wu, Wang, Duan, and Brown}}]{trout2018simulating}
\bibinfo{author}{\bibfnamefont{C.~J.} \bibnamefont{Trout}},
  \bibinfo{author}{\bibfnamefont{M.}~\bibnamefont{Li}},
  \bibinfo{author}{\bibfnamefont{M.}~\bibnamefont{Guti{\'e}rrez}},
  \bibinfo{author}{\bibfnamefont{Y.}~\bibnamefont{Wu}},
  \bibinfo{author}{\bibfnamefont{S.-T.} \bibnamefont{Wang}},
  \bibinfo{author}{\bibfnamefont{L.}~\bibnamefont{Duan}}, \bibnamefont{and}
  \bibinfo{author}{\bibfnamefont{K.~R.} \bibnamefont{Brown}},
  \bibinfo{journal}{New Journal of Physics} \textbf{\bibinfo{volume}{20}},
  \bibinfo{pages}{043038} (\bibinfo{year}{2018}).

\bibitem[{\citenamefont{Wang et~al.}(2009)\citenamefont{Wang, Fowler, Stephens,
  and Hollenberg}}]{Wang:2009}
\bibinfo{author}{\bibfnamefont{D.~S.} \bibnamefont{Wang}},
  \bibinfo{author}{\bibfnamefont{A.~G.} \bibnamefont{Fowler}},
  \bibinfo{author}{\bibfnamefont{A.~M.} \bibnamefont{Stephens}},
  \bibnamefont{and} \bibinfo{author}{\bibfnamefont{L.~C.~L.}
  \bibnamefont{Hollenberg}}, \bibinfo{journal}{arXiv preprint arXiv:0905.0531}
  (\bibinfo{year}{2009}).

\bibitem[{\citenamefont{Bacon}(2006)}]{Bacon:2006}
\bibinfo{author}{\bibfnamefont{D.}~\bibnamefont{Bacon}},
  \bibinfo{journal}{Physical Review A} \textbf{\bibinfo{volume}{73}},
  \bibinfo{pages}{012340} (\bibinfo{year}{2006}).

\bibitem[{\citenamefont{Aliferis and Cross}(2007)}]{Aliferis:2007}
\bibinfo{author}{\bibfnamefont{P.}~\bibnamefont{Aliferis}} \bibnamefont{and}
  \bibinfo{author}{\bibfnamefont{A.~W.} \bibnamefont{Cross}},
  \bibinfo{journal}{Physical Review Letters} \textbf{\bibinfo{volume}{98}},
  \bibinfo{pages}{220502} (\bibinfo{year}{2007}).

\bibitem[{\citenamefont{Li et~al.}(2018)\citenamefont{Li, Miller, and
  Brown}}]{Li:2018}
\bibinfo{author}{\bibfnamefont{M.}~\bibnamefont{Li}},
  \bibinfo{author}{\bibfnamefont{D.}~\bibnamefont{Miller}}, \bibnamefont{and}
  \bibinfo{author}{\bibfnamefont{K.~R.} \bibnamefont{Brown}},
  \bibinfo{journal}{arXiv preprint arXiv:1804.01127}  (\bibinfo{year}{2018}).

\bibitem[{\citenamefont{Yoder}(2017)}]{Yoder:2017}
\bibinfo{author}{\bibfnamefont{T.~J.} \bibnamefont{Yoder}},
  \bibinfo{journal}{arXiv preprint arXiv:1705.01686}  (\bibinfo{year}{2017}).

\bibitem[{\citenamefont{Pastawski et~al.}(2009)\citenamefont{Pastawski, Kay,
  Schuch, and Cirac}}]{Pastawski:2009}
\bibinfo{author}{\bibfnamefont{F.}~\bibnamefont{Pastawski}},
  \bibinfo{author}{\bibfnamefont{A.}~\bibnamefont{Kay}},
  \bibinfo{author}{\bibfnamefont{N.}~\bibnamefont{Schuch}}, \bibnamefont{and}
  \bibinfo{author}{\bibfnamefont{I.}~\bibnamefont{Cirac}},
  \bibinfo{journal}{arXiv preprint arXiv:0911.3843}  (\bibinfo{year}{2009}).

\bibitem[{\citenamefont{Nussinov and Van
  Den~Brink}(2015)}]{nussinov2015compass}
\bibinfo{author}{\bibfnamefont{Z.}~\bibnamefont{Nussinov}} \bibnamefont{and}
  \bibinfo{author}{\bibfnamefont{J.}~\bibnamefont{Van Den~Brink}},
  \bibinfo{journal}{Reviews of Modern Physics} \textbf{\bibinfo{volume}{87}},
  \bibinfo{pages}{1} (\bibinfo{year}{2015}).

\bibitem[{\citenamefont{Delfosse
  et~al.}(2016{\natexlab{a}})\citenamefont{Delfosse, Iyer, and
  Poulin}}]{delfosse2016generalized}
\bibinfo{author}{\bibfnamefont{N.}~\bibnamefont{Delfosse}},
  \bibinfo{author}{\bibfnamefont{P.}~\bibnamefont{Iyer}}, \bibnamefont{and}
  \bibinfo{author}{\bibfnamefont{D.}~\bibnamefont{Poulin}},
  \bibinfo{journal}{arXiv preprint arXiv:1606.07116}
  (\bibinfo{year}{2016}{\natexlab{a}}).

\bibitem[{\citenamefont{Aliferis and Preskill}(2008)}]{PhysRevA.78.052331}
\bibinfo{author}{\bibfnamefont{P.}~\bibnamefont{Aliferis}} \bibnamefont{and}
  \bibinfo{author}{\bibfnamefont{J.}~\bibnamefont{Preskill}},
  \bibinfo{journal}{Physical Review A} \textbf{\bibinfo{volume}{78}},
  \bibinfo{pages}{052331} (\bibinfo{year}{2008}).

\bibitem[{\citenamefont{Webster et~al.}(2015)\citenamefont{Webster, Bartlett,
  and Poulin}}]{PhysRevA.92.062309}
\bibinfo{author}{\bibfnamefont{P.}~\bibnamefont{Webster}},
  \bibinfo{author}{\bibfnamefont{S.~D.} \bibnamefont{Bartlett}},
  \bibnamefont{and} \bibinfo{author}{\bibfnamefont{D.}~\bibnamefont{Poulin}},
  \bibinfo{journal}{Physical Review A} \textbf{\bibinfo{volume}{92}},
  \bibinfo{pages}{062309} (\bibinfo{year}{2015}).

\bibitem[{\citenamefont{Stephens et~al.}(2013)\citenamefont{Stephens, Munro,
  and Nemoto}}]{stephens2013high}
\bibinfo{author}{\bibfnamefont{A.~M.} \bibnamefont{Stephens}},
  \bibinfo{author}{\bibfnamefont{W.~J.} \bibnamefont{Munro}}, \bibnamefont{and}
  \bibinfo{author}{\bibfnamefont{K.}~\bibnamefont{Nemoto}},
  \bibinfo{journal}{Physical Review A} \textbf{\bibinfo{volume}{88}},
  \bibinfo{pages}{060301} (\bibinfo{year}{2013}).

\bibitem[{\citenamefont{Tuckett et~al.}(2018)\citenamefont{Tuckett, Bartlett,
  and Flammia}}]{Tuckett:2018}
\bibinfo{author}{\bibfnamefont{D.~K.} \bibnamefont{Tuckett}},
  \bibinfo{author}{\bibfnamefont{S.~D.} \bibnamefont{Bartlett}},
  \bibnamefont{and} \bibinfo{author}{\bibfnamefont{S.~T.}
  \bibnamefont{Flammia}}, \bibinfo{journal}{Physical review letters}
  \textbf{\bibinfo{volume}{120}}, \bibinfo{pages}{050505}
  (\bibinfo{year}{2018}).

\bibitem[{\citenamefont{Delfosse and Tillich}(2014)}]{delfosse2014decoding}
\bibinfo{author}{\bibfnamefont{N.}~\bibnamefont{Delfosse}} \bibnamefont{and}
  \bibinfo{author}{\bibfnamefont{J.-P.} \bibnamefont{Tillich}}, in
  \emph{\bibinfo{booktitle}{Information Theory (ISIT), 2014 IEEE International
  Symposium on}} (\bibinfo{organization}{IEEE}, \bibinfo{year}{2014}), pp.
  \bibinfo{pages}{1071--1075}.

\bibitem[{\citenamefont{Nickerson and Brown}(2017)}]{nickerson2017analysing}
\bibinfo{author}{\bibfnamefont{N.~H.} \bibnamefont{Nickerson}}
  \bibnamefont{and} \bibinfo{author}{\bibfnamefont{B.~J.} \bibnamefont{Brown}},
  \bibinfo{journal}{arXiv preprint arXiv:1712.00502}  (\bibinfo{year}{2017}).

\bibitem[{\citenamefont{Darmawan and Poulin}(2018)}]{darmawan2018efficient}
\bibinfo{author}{\bibfnamefont{A.~S.} \bibnamefont{Darmawan}} \bibnamefont{and}
  \bibinfo{author}{\bibfnamefont{D.}~\bibnamefont{Poulin}},
  \bibinfo{journal}{arXiv preprint arXiv:1801.01879}  (\bibinfo{year}{2018}).

\bibitem[{\citenamefont{Robertson et~al.}(2017)\citenamefont{Robertson,
  Granade, Bartlett, and Flammia}}]{PhysRevApplied.8.064004}
\bibinfo{author}{\bibfnamefont{A.}~\bibnamefont{Robertson}},
  \bibinfo{author}{\bibfnamefont{C.}~\bibnamefont{Granade}},
  \bibinfo{author}{\bibfnamefont{S.~D.} \bibnamefont{Bartlett}},
  \bibnamefont{and} \bibinfo{author}{\bibfnamefont{S.~T.}
  \bibnamefont{Flammia}}, \bibinfo{journal}{Physical Review Applied}
  \textbf{\bibinfo{volume}{8}}, \bibinfo{pages}{064004} (\bibinfo{year}{2017}).

\bibitem[{\citenamefont{Napp and Preskill}(2012)}]{Napp:2012}
\bibinfo{author}{\bibfnamefont{J.}~\bibnamefont{Napp}} \bibnamefont{and}
  \bibinfo{author}{\bibfnamefont{J.}~\bibnamefont{Preskill}},
  \bibinfo{journal}{arXiv preprint arXiv:1209.0794}  (\bibinfo{year}{2012}).

\bibitem[{\citenamefont{Brooks and Preskill}(2013)}]{PhysRevA.87.032310}
\bibinfo{author}{\bibfnamefont{P.}~\bibnamefont{Brooks}} \bibnamefont{and}
  \bibinfo{author}{\bibfnamefont{J.}~\bibnamefont{Preskill}},
  \bibinfo{journal}{Physical Review A} \textbf{\bibinfo{volume}{87}},
  \bibinfo{pages}{032310} (\bibinfo{year}{2013}).

\bibitem[{\citenamefont{Aliferis et~al.}(2009)\citenamefont{Aliferis, Brito,
  DiVincenzo, Preskill, Steffen, and Terhal}}]{aliferis2009fault}
\bibinfo{author}{\bibfnamefont{P.}~\bibnamefont{Aliferis}},
  \bibinfo{author}{\bibfnamefont{F.}~\bibnamefont{Brito}},
  \bibinfo{author}{\bibfnamefont{D.~P.} \bibnamefont{DiVincenzo}},
  \bibinfo{author}{\bibfnamefont{J.}~\bibnamefont{Preskill}},
  \bibinfo{author}{\bibfnamefont{M.}~\bibnamefont{Steffen}}, \bibnamefont{and}
  \bibinfo{author}{\bibfnamefont{B.~M.} \bibnamefont{Terhal}},
  \bibinfo{journal}{New Journal of Physics} \textbf{\bibinfo{volume}{11}},
  \bibinfo{pages}{013061} (\bibinfo{year}{2009}).

\bibitem[{\citenamefont{Delfosse
  et~al.}(2016{\natexlab{b}})\citenamefont{Delfosse, Iyer, and
  Poulin}}]{delfosse2016linear}
\bibinfo{author}{\bibfnamefont{N.}~\bibnamefont{Delfosse}},
  \bibinfo{author}{\bibfnamefont{P.}~\bibnamefont{Iyer}}, \bibnamefont{and}
  \bibinfo{author}{\bibfnamefont{D.}~\bibnamefont{Poulin}},
  \bibinfo{journal}{arXiv preprint arXiv:1611.04256}
  (\bibinfo{year}{2016}{\natexlab{b}}).

\bibitem[{\citenamefont{Bomb{\'\i}n}(2010)}]{Bombin:2010b}
\bibinfo{author}{\bibfnamefont{H.}~\bibnamefont{Bomb{\'\i}n}},
  \bibinfo{journal}{Physical Review A} \textbf{\bibinfo{volume}{81}},
  \bibinfo{pages}{032301} (\bibinfo{year}{2010}).

\bibitem[{\citenamefont{Delfosse and Nickerson}(2017)}]{delfosse2017almost}
\bibinfo{author}{\bibfnamefont{N.}~\bibnamefont{Delfosse}} \bibnamefont{and}
  \bibinfo{author}{\bibfnamefont{N.~H.} \bibnamefont{Nickerson}},
  \bibinfo{journal}{arXiv preprint arXiv:1709.06218}  (\bibinfo{year}{2017}).

\bibitem[{\citenamefont{Dorier et~al.}(2005)\citenamefont{Dorier, Becca, and
  Mila}}]{dorier2005quantum}
\bibinfo{author}{\bibfnamefont{J.}~\bibnamefont{Dorier}},
  \bibinfo{author}{\bibfnamefont{F.}~\bibnamefont{Becca}}, \bibnamefont{and}
  \bibinfo{author}{\bibfnamefont{F.}~\bibnamefont{Mila}},
  \bibinfo{journal}{Physical Review B} \textbf{\bibinfo{volume}{72}},
  \bibinfo{pages}{024448} (\bibinfo{year}{2005}).

\bibitem[{\citenamefont{Kugel and Khomskii}(1973)}]{kugel1973crystal}
\bibinfo{author}{\bibfnamefont{K.}~\bibnamefont{Kugel}} \bibnamefont{and}
  \bibinfo{author}{\bibfnamefont{D.}~\bibnamefont{Khomskii}},
  \bibinfo{journal}{Zh. {\'E}ksp. Teor. Fiz} \textbf{\bibinfo{volume}{64}},
  \bibinfo{pages}{1429} (\bibinfo{year}{1973}).

\bibitem[{\citenamefont{Paetznick and Reichardt}(2013)}]{Paetznick:2013}
\bibinfo{author}{\bibfnamefont{A.}~\bibnamefont{Paetznick}} \bibnamefont{and}
  \bibinfo{author}{\bibfnamefont{B.~W.} \bibnamefont{Reichardt}},
  \bibinfo{journal}{Physical review letters} \textbf{\bibinfo{volume}{111}},
  \bibinfo{pages}{090505} (\bibinfo{year}{2013}).

\bibitem[{\citenamefont{Calderbank and Shor}(1996)}]{calderbank1996good}
\bibinfo{author}{\bibfnamefont{A.~R.} \bibnamefont{Calderbank}}
  \bibnamefont{and} \bibinfo{author}{\bibfnamefont{P.~W.} \bibnamefont{Shor}},
  \bibinfo{journal}{Physical Review A} \textbf{\bibinfo{volume}{54}},
  \bibinfo{pages}{1098} (\bibinfo{year}{1996}).

\bibitem[{\citenamefont{Steane}(1996)}]{steane1996error}
\bibinfo{author}{\bibfnamefont{A.~M.} \bibnamefont{Steane}},
  \bibinfo{journal}{Physical Review Letters} \textbf{\bibinfo{volume}{77}},
  \bibinfo{pages}{793} (\bibinfo{year}{1996}).

\bibitem[{\citenamefont{Kubica et~al.}(2018)\citenamefont{Kubica, Beverland,
  Brand{\~a}o, Preskill, and Svore}}]{kubica2018three}
\bibinfo{author}{\bibfnamefont{A.}~\bibnamefont{Kubica}},
  \bibinfo{author}{\bibfnamefont{M.~E.} \bibnamefont{Beverland}},
  \bibinfo{author}{\bibfnamefont{F.}~\bibnamefont{Brand{\~a}o}},
  \bibinfo{author}{\bibfnamefont{J.}~\bibnamefont{Preskill}}, \bibnamefont{and}
  \bibinfo{author}{\bibfnamefont{K.~M.} \bibnamefont{Svore}},
  \bibinfo{journal}{Physical Review Letters} \textbf{\bibinfo{volume}{120}},
  \bibinfo{pages}{180501} (\bibinfo{year}{2018}).

\bibitem[{\citenamefont{Edmonds}(2009)}]{edmonds2009paths}
\bibinfo{author}{\bibfnamefont{J.}~\bibnamefont{Edmonds}}, in
  \emph{\bibinfo{booktitle}{Classic Papers in Combinatorics}}
  (\bibinfo{publisher}{Springer}, \bibinfo{year}{2009}), pp.
  \bibinfo{pages}{361--379}.

\bibitem[{\citenamefont{Fowler}(2013)}]{fowler2013accurate}
\bibinfo{author}{\bibfnamefont{A.~G.} \bibnamefont{Fowler}},
  \bibinfo{journal}{Physical Review A} \textbf{\bibinfo{volume}{87}},
  \bibinfo{pages}{062320} (\bibinfo{year}{2013}).

\bibitem[{\citenamefont{Kruskal}(1956)}]{kruskal1956shortest}
\bibinfo{author}{\bibfnamefont{J.~B.} \bibnamefont{Kruskal}},
  \bibinfo{journal}{Proceedings of the American Mathematical society}
  \textbf{\bibinfo{volume}{7}}, \bibinfo{pages}{48} (\bibinfo{year}{1956}).

\bibitem[{\citenamefont{Delfosse and Z{\'e}mor}(2017)}]{delfosse2017linear}
\bibinfo{author}{\bibfnamefont{N.}~\bibnamefont{Delfosse}} \bibnamefont{and}
  \bibinfo{author}{\bibfnamefont{G.}~\bibnamefont{Z{\'e}mor}},
  \bibinfo{journal}{arXiv preprint arXiv:1703.01517}  (\bibinfo{year}{2017}).

\bibitem[{\citenamefont{Nishimori}(1986)}]{nishimori1986geometry}
\bibinfo{author}{\bibfnamefont{H.}~\bibnamefont{Nishimori}},
  \bibinfo{journal}{Journal of the Physical Society of Japan}
  \textbf{\bibinfo{volume}{55}}, \bibinfo{pages}{3305} (\bibinfo{year}{1986}).

\bibitem[{\citenamefont{Dotsenko et~al.}(1991)\citenamefont{Dotsenko, Selke,
  and Talapov}}]{cluster}
\bibinfo{author}{\bibfnamefont{V.~S.} \bibnamefont{Dotsenko}},
  \bibinfo{author}{\bibfnamefont{W.}~\bibnamefont{Selke}}, \bibnamefont{and}
  \bibinfo{author}{\bibfnamefont{A.}~\bibnamefont{Talapov}},
  \bibinfo{journal}{Physica A: Statistical Mechanics and its Applications}
  \textbf{\bibinfo{volume}{170}}, \bibinfo{pages}{278} (\bibinfo{year}{1991}).

\bibitem[{\citenamefont{Binder}(1981)}]{Binder1981}
\bibinfo{author}{\bibfnamefont{K.}~\bibnamefont{Binder}},
  \bibinfo{journal}{Zeitschrift f{\"u}r Physik B Condensed Matter}
  \textbf{\bibinfo{volume}{43}}, \bibinfo{pages}{119} (\bibinfo{year}{1981}).

\bibitem[{\citenamefont{Honecker et~al.}(2001)\citenamefont{Honecker, Picco,
  and Pujol}}]{Honecker:2000}
\bibinfo{author}{\bibfnamefont{A.}~\bibnamefont{Honecker}},
  \bibinfo{author}{\bibfnamefont{M.}~\bibnamefont{Picco}}, \bibnamefont{and}
  \bibinfo{author}{\bibfnamefont{P.}~\bibnamefont{Pujol}},
  \bibinfo{journal}{Physical Review Letters} \textbf{\bibinfo{volume}{87}},
  \bibinfo{pages}{047201} (\bibinfo{year}{2001}).

\bibitem[{\citenamefont{Shor}(1995)}]{Shor:1995b}
\bibinfo{author}{\bibfnamefont{P.~W.} \bibnamefont{Shor}},
  \bibinfo{journal}{Physical Review A} \textbf{\bibinfo{volume}{52}},
  \bibinfo{pages}{R2493} (\bibinfo{year}{1995}).

\bibitem[{\citenamefont{Fujii and Tokunaga}(2012)}]{fujii2012error}
\bibinfo{author}{\bibfnamefont{K.}~\bibnamefont{Fujii}} \bibnamefont{and}
  \bibinfo{author}{\bibfnamefont{Y.}~\bibnamefont{Tokunaga}},
  \bibinfo{journal}{Physical Review A} \textbf{\bibinfo{volume}{86}},
  \bibinfo{pages}{020303} (\bibinfo{year}{2012}).

\bibitem[{\citenamefont{R{\"o}thlisberger
  et~al.}(2012)\citenamefont{R{\"o}thlisberger, Wootton, Heath, Pachos, and
  Loss}}]{rothlisberger2012incoherent}
\bibinfo{author}{\bibfnamefont{B.}~\bibnamefont{R{\"o}thlisberger}},
  \bibinfo{author}{\bibfnamefont{J.~R.} \bibnamefont{Wootton}},
  \bibinfo{author}{\bibfnamefont{R.~M.} \bibnamefont{Heath}},
  \bibinfo{author}{\bibfnamefont{J.~K.} \bibnamefont{Pachos}},
  \bibnamefont{and} \bibinfo{author}{\bibfnamefont{D.}~\bibnamefont{Loss}},
  \bibinfo{journal}{Physical Review A} \textbf{\bibinfo{volume}{85}},
  \bibinfo{pages}{022313} (\bibinfo{year}{2012}).

\bibitem[{\citenamefont{Huang and Brown}(2018)}]{Huang:2018}
\bibinfo{author}{\bibfnamefont{S.}~\bibnamefont{Huang}} \bibnamefont{and}
  \bibinfo{author}{\bibfnamefont{K.~R.} \bibnamefont{Brown}},
  \bibinfo{journal}{In preparation}  (\bibinfo{year}{2018}).

\bibitem[{\citenamefont{Maskara et~al.}(2018)\citenamefont{Maskara, Kubica, and
  Jochym-O'Connor}}]{maskara2018advantages}
\bibinfo{author}{\bibfnamefont{N.}~\bibnamefont{Maskara}},
  \bibinfo{author}{\bibfnamefont{A.}~\bibnamefont{Kubica}}, \bibnamefont{and}
  \bibinfo{author}{\bibfnamefont{T.}~\bibnamefont{Jochym-O'Connor}},
  \bibinfo{journal}{arXiv preprint arXiv:1802.08680}  (\bibinfo{year}{2018}).

\end{thebibliography}

\end{document}